\newif\ifcomment
\commenttrue

\ifcomment
    \newcounter{MVNumberOfComments}
    \stepcounter{MVNumberOfComments}
    \newcommand{\mvnote}[1]{\textcolor{blue}{\small \bf [MV\#\arabic{MVNumberOfComments}\stepcounter{MVNumberOfComments}: #1]}}

    \newcounter{HCNumberOfComments}
    \stepcounter{HCNumberOfComments}
    \newcommand{\hcnote}[1]{\textcolor{brown}{\small \bf [HC\#\arabic{HCNumberOfComments}\stepcounter{HCNumberOfComments}: #1]}}
    
    \newcounter{YZNumberOfComments}
    \stepcounter{YZNumberOfComments}

    \newcounter{ARNumberOfComments}
    \stepcounter{ARNumberOfComments}
    \newcommand{\arnote}[1]{\textcolor{red}{\small \bf [AR\#\arabic{ARNumberOfComments}\stepcounter{ARNumberOfComments}: #1]}}

    \newcounter{NSNumberOfComments}
    \stepcounter{NSNumberOfComments}
    \newcommand{\nsnote}[1]{\textcolor{orange}{\small \bf [NS\#\arabic{NSNumberOfComments}\stepcounter{NSNumberOfComments}: #1]}}
    
    \newcommand{\NOTE}[1]
    {
      {\footnotesize\it
        \begin{center}
          \begin{tabular}{|c|}
           \hline
            \parbox{0.85\columnwidth}{
              \medskip
              #1
              \medskip} \\
            \hline
          \end{tabular}
        \end{center}
        }
    }

\else
    \newcommand\mvnote[1]{}
    \newcommand\arnote[1]{}
    \newcommand\hcnote[1]{}
    \newcommand\nsnote[1]{}
    \newcommand\NOTE[1]{}
\fi

\newcommand{\eg}{{e.g.,}\xspace}

\newcommand{\ie}{{\it i.e.,}\xspace}
\newcommand{\folder}{./fig}

\newcommand{\one}{\emph{(i)}\xspace}
\newcommand{\two}{\emph{(ii)}\xspace}

\newcommand{\ripe}{{RIPE Atlas}\xspace}

\newcounter{NumTakeaways}
\stepcounter{NumTakeaways}

\newcommand{\pb}[1]{\vspace{0.5ex}\noindent{\textbf{\emph{#1}\hspace*{.3em}}}}

\documentclass[10pt,twocolumn]{article}
\usepackage[a4paper]{geometry}
\usepackage{xspace}
\usepackage{multirow}
\usepackage{subfig} 
\usepackage[most]{tcolorbox}
\PassOptionsToPackage{hyphens}{url}\usepackage{hyperref}

\usepackage{tikz}
\newcommand*\circled[1]{\tikz[baseline=(char.base)]{
    \node[font=\footnotesize,align=center,shape=circle,text=white,fill=black,inner sep=0.1pt] (char) {#1};}}

\newcommand{\camfix}[1]{\textcolor{blue}{#1}}
\newcommand{\reviewfix}[1]{\textsuperscript{\textcolor{red}{\textit{#1}}}}

\renewcommand{\camfix}[1]{\textcolor{black}{#1}}
\renewcommand{\reviewfix}[1]{}

\date{}

\begin{document}
\title{\textbf{\Large Dissecting the Performance of  Satellite Network Operators\let\thefootnote\relax\footnote{Corresponding author: Aravindh Raman (aravindh.raman@gmail.com)}}}
\author{Aravindh Raman\\ \small Telefonica Research
\and Matteo Varvello\\ \small Nokia Bell Labs
\and Hyunseok Chang\\ \small Nokia Bell Labs 
\and Nishanth Sastry\\ \small University of Surrey \and Yasir Zaki\\ \small NYU Abu Dhabi
}

\maketitle

\begin{abstract}
  The rapid growth of satellite network operators (SNOs) has revolutionized broadband communications, enabling global connectivity and bridging the digital divide. As these networks expand, it is important to evaluate their performance and efficiency. This paper presents the first comprehensive study of SNOs. We take an opportunistic approach and devise a methodology which allows to identify public network measurements performed via SNOs. We apply this methodology to both M-Lab and RIPE public datasets which allowed us to characterize low level performance and footprint of up to 18 SNOs operating in different orbits. Finally, we identify and recruit paid testers on three popular SNOs (Starlink, HughesNet, and Viasat) to evaluate the performance of popular applications like web browsing and video streaming. 
\end{abstract}

\maketitle

\section{Introduction}
In recent years, satellite network operators (SNOs) have gained significant attention as an alternative to terrestrial Internet, especially in remote or underserved areas. Companies like Starlink~\cite{starlink}, Viasat~\cite{viasat}, and HughesNet~\cite{hughesnet} have invested heavily in satellite technologies to provide high-speed Internet access to their customers regardless of their location. Satellite technologies have evolved from using mainly geosynchronous (GEO) satellites -- which operate at around 35,000~km above the earth surface~\cite{maral2020satellite} -- to Low Earth Orbit (LEO) satellites --- which can be as close as 550~km away as in the case of the closest orbital shell from Starlink~\cite{Cakaj2021starlink}.

Several research papers~\cite{kassem2022browser,leo22,michel22} have investigated the performance of SNOs, especially focusing on Starlink due to its recent popularity and novel LEO-based technology. The main challenge in performing such measurements is gathering vantage points, \ie instrumentable devices from where to perform network measurements. To tackle this challenge, researchers have deployed a few controlled nodes with dedicated satellite links (from one to four~\cite{michel22,leo22}) or recruited Starlink users to perform network measurements via a browser addon~\cite{kassem2022browser}.

The main goal of this study is to \textit{scale up} the previous measurements of SNOs in multiple dimensions. More specifically, we aim to expand the scope of the measurements in terms of \textit{space} (\ie more geographic locations), \textit{time} (\ie longer measurement periods), \textit{satellite technologies} (\ie multiple SNOs), and \textit{application diversity} (\ie multiple applications). In this type of large-scale measurement study, the main challenge is data collection. Indeed, building our own measurement testbed from scratch is challenging, and hard to scale. Not to mention that it would take years to deploy such a testbed, during which lots of data would be lost over time.

Motivated by the above, we opt for an \emph{opportunistic} approach and leverage data and measurement endpoints already available. Our approach is to \textit{identify} SNO measurements from public datasets (M-Lab~\cite{mlab} and Ripe Atlas~\cite{ripeatlas}), and \textit{recruit} SNO users from crowdsourcing platforms (Prolific~\cite{prolific}). Public measurements have the advantage of their scale, \ie covering many SNOs over a long period of time, but they are limited to low-level measurements, such as \texttt{traceroute} and speed tests. Actual SNO users can run more complex measurements, such as Web browsing and video streaming, but likely at a smaller scale in space and time.

For all these different data sources, we first need to correctly identify measurements and testers associated with SNOs. This is challenging since very little ground truth is available, and SNOs are complex entities offering mixed connectivity, \eg both LEO and GEO as well as satellite and wireline access. To tackle this challenge, we devise the following methodology. First, we leverage ASNs and public information to identify \textit{potential} SNOs along with their access technology (LEO, MEO, or GEO). Next, we use public data from M-Lab to build maps of network latencies per ASN and IP blocks with an AS. We then filter ASNs and IP blocks whose latency profiles show even the slightest incompatibility with the satellite technology offered by their SNO, \eg less than 500 ms over GEO.  This strict filtering identifies 10 SNOs in M-Lab public dataset, and also produces accurate latency profiles per SNO access technology. We then use this information to relax the previous filtering and identify data associated with an additional 8 SNOs, for a total of 18 SNOs.

After applying the above SNO detection mechanism, we identify \camfix{11.92}\reviewfix{D2} million TCP-based speed tests from 18 SNOs (2 LEO, 1 Medium Earth Orbit or MEO, and 15 GEO) in the M-Lab dataset (between January 2021 and March 2023), and about 6 million \texttt{traceroute} measurements from 67 probes connected via Starlink (LEO) in the \ripe dataset (between May 2022 and May 2023). For Prolific, we find that their prescreening APIs are only partially effective in identifying SNO subscribers. We instead run a \textit{census} which identifies, out of 14,371 participants, 57 potential testers connected via Starlink, HughesNet, and Viasat.  Over one month, we recruited 20 out of these 57 testers willing to install and run a browser addon we developed to measure application performance. In the following, we summarize our key findings.

\pb{LEO vs.~MEO vs.~GEO.} Our large dataset allows us to perform what we believe is the first large-scale and global characterization of the different SNO technologies available. As expected, we find that LEO supports much higher throughput and much lower latencies than GEO or MEO. However, we also find interesting patterns -- for instance, LEO networks suffer from much higher jitter variation, likely due to frequent satellite handoffs. This may negatively affect applications that expect a consistent latency profile. On the other hand, customers of most GEO networks suffer a high amount of data retransmissions, likely due to transport level retransmission timeouts. Fortunately, Performance Enhancing Proxies (PEPs)~\cite{rfc2488, rfc3449} appear to mitigate this problem and GEO operators which employ PEPs have retransmission profiles similar to LEO operators.

\pb{PoP selection matters.} On average, Starlink LEO satellites provide connectivity to their customers while only adding an extra 30-40~ms. To achieve such low latency, proper Point-of-Presence (PoP) selection is paramount. We notably identify two main examples of this observation. First, as of today, Starlink customers in the Philippines experience 2x latencies (80~ms) because they are associated to a PoP in Tokyo, Japan (``customer.tkyojpn1.pop.starlinkisp.net''). Second, New Zealand Starlink customers experienced a 20~ms latency reduction since July 2022 when their PoP changed from Sydney (Australia) to Auckland (New Zealand).

\pb{Careful technology selection.} Modern networking technologies are important for users in GEO-based SNOs like HughesNet and Viasat. Our experiments show that selecting the fastest content-delivery network (CDN), Fastly in our experiments, can reduce download time by up to one second, when loading critical Web component like popular JavaScript (JS) libraries. A similar effect is achieved by JS \textit{minification}, or the process to minimize JS code. Recent evolution of the HTTP protocol, like HTTP/2, also allows to bridge the gap between GEO and LEO users on HTTP/1.1, thanks to features such as connection multiplexing, which result in a reduction in the number of (secure) connections and allow efficient usage of the underlying transport.

\section{Background and Related Work}
\label{sec:background}

\pb{Internet measurement platforms.}
There are several prominent open platforms for Internet measurements. We here provide more details about the two platforms we leverage in this paper: M-Lab~\cite{mlab} and RIPE Atlas~\cite{ripeatlas}. M-Lab allows researchers to deploy web-based Internet measurement tools to be run by end users in the wild. It operates server pods worldwide, each containing three to four interconnected servers. M-Lab makes the measurement data collected from end users publicly accessible via bulk downloads or BigQuery~\cite{mlabbigquery}. RIPE Atlas is a distributed measurement platform powered by thousands of dedicated measurement devices called ``probes'' hosted by volunteers all around the world.  Each probe periodically performs a set of ``built-in'' measurements: ping, traceroute, DNS, SSL/TLS and HTTP probing. These measurements are primarily directed towards well-known targets, such as DNS root servers and part of the RIPE Atlas infrastructure~\cite{ripemeasure}.  Similar to M-Lab, RIPE Atlas makes collected measurement data publicly available via  BigQuery.

Datasets from these platforms have enabled a wide range of measurement studies, \eg assessing the impact of major societal events~\cite{jain2022ukranian,candela20}, characterizing access networks~\cite{bajpai17,fries22}, inferring global or regional Internet properties~\cite{khalid19,candela19}, uncovering network anomalies~\cite{ben16,hou21}, evaluating cloud reachability~\cite{corneo21}, etc. To the best of our knowledge, however, no prior study has analyzed the data sets with a focus on global SNOs.

\pb{Satellite Internet measurements.}
Kassem et al.~\cite{kassem2022browser} studied Starlink connectivity by utilizing a combination of a custom browser extension and dedicated measurement nodes. They observed non-negligible variability in web access performance across different weather conditions and geographic locations, and significant packet loss rates as high as 50\%. These findings are based on browser extension data collected from 28 users in 10 cities worldwide and three dedicated nodes provisioned in UK, Spain and USA. \camfix{The study also compares the performance of traditional access networks (cellular and fiber) with that of LEO-based access.}\reviewfix{C2} Ma et al.~\cite{leo22} carried out a measurement study on Starlink by utilizing four dedicated Starlink kits stationed over a range of terrains including major cities and remote areas in west Canada.  In areas where terrestrial broadband access was available, they also compared Starlink access against terrestrial Internet access. They reported similar observations where Starlink connectivity is subject to more dynamic throughput and latency variation than terrestrial networks and is heavily affected by environments such as terrain characteristics and weather conditions. The authors of \cite{perdices2022satellite} performed a detailed study on the traffic carried by a GEO-based SatCom network. By exploiting the fact that all subscriber traffic is relayed via the central ground station, they deployed a traffic monitoring server at the ground station. Their study compares subscriber access patterns across different geographic regions, characterizes RTT, DNS and download performance, and highlights technical challenges faced by the SatCom infrastructure. Finally, several works~\cite{claypool20,endres22,netsys19,michel22} studied congestion control behaviors and performance of various transport protocols such as TCP, HTTP and QUIC on satellite Internet. These studies relied on a single vantage point, \ie a controlled lab environment with dedicated satellite links. 

Our paper differs from related works in two key aspects. First, it devises a methodology to identify, from public datasets, experiments run via SNOs. Second, it compares the performance of up to 18 SNOs from multiple different angles (orbit of operations, application, terrestrial connectivity, etc.). \looseness=-1

\section{Methodology}
\label{sec:meth}
This section describes the methodology we have devised to dissect the performance of global SNOs. Our approach is to \textit{identify} SNO measurements from public datasets (see Section~\ref{sec:meth:datasets}), and \textit{recruit} SNO users from crowdsourcing platforms (see Section~\ref{sec:meth:prolific}). %
In both approaches, the first step is to correctly identify the measurements or users that are associated with SNOs (see Section~\ref{sec:meth:identify}). 

\subsection{Public Datasets}
\label{sec:meth:datasets}
\pb{M-Lab's Speed Test.} This dataset provides upload and download speeds measured by clients distributed across the world using the Network Diagnostic Tool (NDT). Currently, NDT exists in two versions, NDT5 and NDT7. However, the majority of clients have migrated to NDT7  (more than 90\% as of 2021~\cite{measurementlab_migrate}). Accordingly, we focus on the more recent NDT7 traces for our analysis. 

An NDT test consists of a single  TCP connection between a client and a nearby M-Lab server determined by Google's location service~\cite{google-geolocation}. The server captures the \texttt{TCP\_Info} for each speed test experiment running a polling loop on every open TCP socket, \camfix{thus capturing multiple records per session}\reviewfix{F4}. The captured trace is uploaded to Google BigQuery, and provides information such as RTT, jitter, delivery rate, and the sizes of sent and received bytes. To overcome a bug in the \texttt{APP\_Info} values reported by the client~\cite{macmillan2023ndt}, we rely on the TCP Info instead. For the latency-based analysis we use 5$^{th}$ percentile value per speed test session. To measure the variability in latencies within a speed test session we measure 95$^{th}$ 
 percentile of jitter and normalize using the 5$^{th}$ percentile latency of that session \ie jitter variability = ${jitter_{p95}/latency_{p5}}$. Appendix~\ref{sec:appendix:ethics} discusses the ethics behind our data collection. \looseness=-1
 
\pb{RIPE Atlas Built-in Measurements.} We gather access to \ripe  between May 3rd, 2022 and May 3rd, 2023.
We  focus on the built-in measurements (see Section~\ref{sec:background}) since, in absence of failures, they are deployed at all probes with a constant frequency. We further focus on \texttt{traceroute} between \ripe probes and the 13 anycast addresses of root DNS servers~\cite{ianaRoot}. We choose \texttt{traceroute} as it provides visibility into Starlink's networks built with multiple PoPs, allowing us to derive round-trip time (RTT) to the Starlink \textit{Carrier Grade NAT gateway} to the Internet (with address ``100.64.0.1''). To derive the physical locations of such PoPs, we perform reverse DNS lookups of the probes' public IP addresses. \camfix{We collect public IP addresses by parsing the `src\_addr' field of the certificates obtained during RIPE ``SSLCert'' built-in measurements~\cite{bajpai2015lessons}. We choose this method instead of extracting them from the probe's metadata because SSLCert experiments are recursive and run every 12 hours, allowing us to capture any temporal changes in the IP addresses. }\reviewfix{E2} %
Reverse DNS lookups map a probe's IP address with a hostname in the form of ``customer.LOCATION.pop.starlinkisp.net'', thus revealing the location of Starlink PoPs.  Each PoP represents a connection point between Starlink and the Internet backbone, serving potentially multiple ``ground stations'' or where LEO satellites connect to. 

While analyzing \ripe dataset, we discovered that some probes were previously hosted by a different ISP than Starlink, and that the probes table only reports the most recent ASN. Further, a few probes are multi-homed and use mobile connections as a ``failover''. To address these issues, we leverage \texttt{traceroute} data to verify that the (private) address of a Starlink Carrier Grade NAT gateway was present on the routing path.

\subsection{Identifying Satellite Network Operators}
\label{sec:meth:identify}

\begin{figure*}
\centering
    \includegraphics[width=0.9\linewidth]{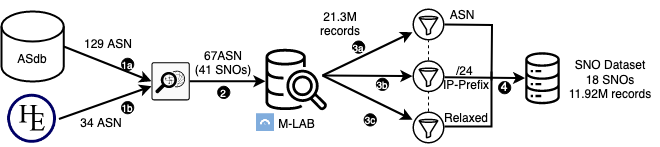}
    \caption{\camfix{Data collection and processing pipeline: Methodology to identify Satellite Network Operators (SNOs) from the M-Lab dataset.}\reviewfix{E3, F1}}
    \label{fig:ds:filter}
    
\end{figure*}

The public datasets we used (M-Lab and \ripe) do not provide reliable information about the characteristics of access links their users or probes are connected to. Similarly, crowdsourcing platforms (\eg Prolific) either do not provide such information or, when they do, it is not always reliable. Intuitively, SNO users and measurements can be identified based on the originating Autonomous System Number (ASN) available either in the measurement datasets or via a tester's IP address. However, previous works have shown that accurate ASN-to-operator mapping information is lacking~\cite{rula2017cellspotting, asdb}. For instance, an SNO might route the wireline traffic of its corporate offices
via the same ASN used for its satellite customers. In addition, SNOs might rely on different technologies (LEO, GEO, MEO, as well as a mix) for which no public ASN-to-technology information is currently available. To tackle this challenge, we devise the following methodology \camfix{(shown in Figure~\ref{fig:ds:filter})}\reviewfix{E3, F1} which we describe below.

\begin{figure*}
\subfloat[\texttt{Starlink}]{\includegraphics[width=0.2\linewidth, trim={0 0 0 0.5cm},clip]{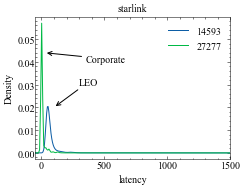}
\label{fig:kde:starlink}} \subfloat[\texttt{OneWeb}]{\includegraphics[width=0.2\linewidth, trim={0 0 0 0.5cm},clip]{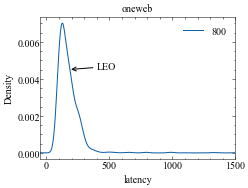}\label{fig:kde:geo}}
 \subfloat[\texttt{O3b}]{\includegraphics[width=0.2\linewidth, trim={0 0 0 0.5cm},clip]{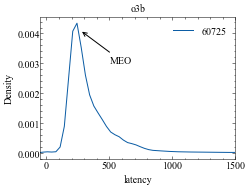}
\label{fig:kde:oneweb}}
\subfloat[\texttt{SES (H)}]{\includegraphics[width=0.2\linewidth, trim={0 0 0 0.5cm},clip]{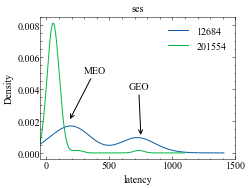}
\label{fig:kde:ses}}
\subfloat[\texttt{TelAlaska (M)}]{\includegraphics[width=0.2\linewidth, trim={0 0 0 0.5cm},clip]{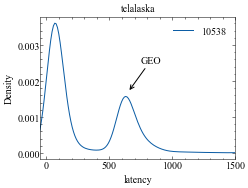}
\label{fig:kde:mixed}}\vspace{-10pt}
\caption{Kernel Density Estimation (KDE) curves by SNO showing standalone ASN for satellite user connections, Hybrid Access (H), Mixed Access (M).}
\vspace{-0.25in}
\label{fig:KDE}
\end{figure*}

\pb{ASN-to-SNO mapping.} \camfix{(steps \circled{1a} and \circled{1b}  in Figure~\ref{fig:ds:filter})} We map a client's ASN to its corresponding operator using \texttt{ASdb}~\cite{asdb} and Hurricane Electric's (HE) BGP toolkit~\cite{hebgp} \reviewfix{E3, F1}. \texttt{ASdb} maps each ASN to its organization and a corresponding category (\eg satellite or fiber) using machine learning approaches driven by manual data curation from various data sources such as Dun \& Bradstreet and web classifiers like Zvelo~\cite{zvleo-blog}. Out of all 105k ASes recorded in \texttt{ASdb}, there are 129 ASes which are associated with the category ``Satellite Communication'' under ``Computer and Information Technology''. However, we observe that several well-known SNOs like Starlink and Viasat are missing from the \texttt{ASdb} dataset. We fill this gap by searching for popular SNOs in the HE's BGP toolkit which reveals an additional 35 ASes. In total, we obtain 164 ASN-to-SNO mappings.

Next, we gather information from IPInfo~\cite{ipinfodevelopers} about these 164 ASNs: organization, IP address ranges, and website. We then visit the website of each ASN present in our datasets (see Section~\ref{sec:meth:datasets}) to augment the ASN-to-SNO mapping with access technology information. In the process, we discover that more than half of the 164 ASes do not necessarily belong to SNOs, instead to other similar providers such as Cable TV operators (\eg Cable Axion), residential broadband (\eg Filer Mutual Telephone), navigation services (\eg Teletrac), teleport operators (\eg United Teleports Inc), etc. After filtering them out manually, we obtain 67 ASNs belonging to 41 SNOs. \camfix{(step \circled{2} in Figure~\ref{fig:ds:filter})}\reviewfix{E3, F1}. Interested readers can refer to Table~\ref{tab:operators} in the Appendix for more information about these SNOs, their associated ASNs. %

\pb{ASN-to-SNO validation.} \camfix{(step \circled{3a} in Figure~\ref{fig:ds:filter})}\reviewfix{E3, F1} For the above 41 SNOs, we extract 21.3 million NDT speed test records from M-Lab traces collected between January, 2021 and April 2023 (see Section~\ref{sec:meth:datasets}). For every NDT speed test, we take the 5$^{th}$ percentile of the  latency estimated via TCP as an indication of the \textit{access latency}. We then plot, for each SNO and ASN, the Kernel Density Estimation (KDE)~\cite{chung_diffusion_2009} --- a statistical technique used to estimate the probability density function of a random variable based on a set of observed data points --- of its access latency. 

Figure~\ref{fig:KDE} shows representative KDE curves for the ASNs associated with Starlink and OneWeb (LEO), O3b (MEO), SES (both MEO and GEO), and TeleAlaska (GEO). The figure shows that LEO operators are characterized by low latencies, although with a significant difference between Starlink (median of 56~ms) and OneWeb (median of 154~ms). Note how comparable MEO latencies (280~ms for O3b and 220~ms for one component of SES) are to those of OneWeb, despite the latter being a LEO operator. GEO latencies, on the other hand, clearly depart from latency distributions of the other technologies, about 700~ms for both SES and TelAlaska. This analysis also unveils clear outliers at the AS level, such as ASN 27277 for Starlink whose KDE curve deviates from expected latencies of a LEO operator. Digging deeper, we find in the PeeringDB notes field of AS14593 (the AS used for Starlink customers) that AS27277 is used for Starlink's corporate network~\cite{peeringdbsl}, \ie its users are on an entirely terrestrial path allowing for lower latencies. Similarly, the KDE curve for ASN 201554 (SES) departs from its expected hybrid (MEO plus GEO) access --  a bimodal distribution following a combination of the respective MEO/GEO KDE curves. The same behavior is also observed within the same ASN, as shown by the low latency peak for the GEO-based TelAlaska~\cite{telalaskaTele}. %

\begin{figure*}
    \subfloat[\texttt{/24 prefix.}]{\includegraphics[width=0.235\linewidth]{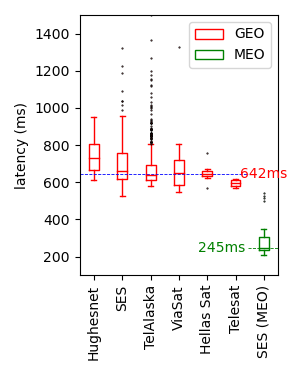}
    \label{fig:ds:prefix}}
    \subfloat[\texttt{Relaxed prefix-based filtering.}]{\includegraphics[width=0.40\linewidth]{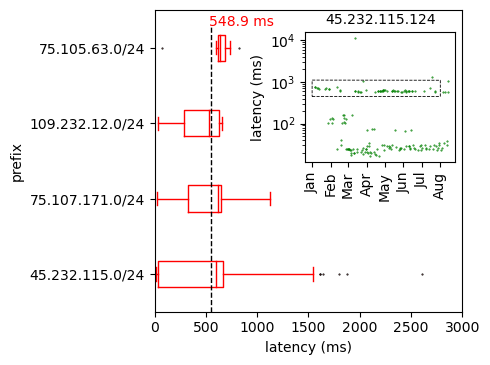}
    \label{fig:ds:lateny}}
    \subfloat[\texttt{RTT across SNOs.}]{\includegraphics[width=0.32  \linewidth]{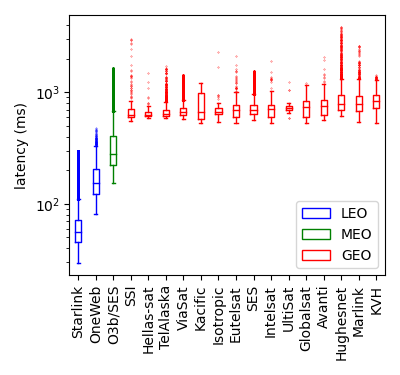}
    \label{fig:sno:rtt}}
    \vspace{-0.1in}
    \caption{Analysis of the methodology we devised to identify SNOs.}
    \label{fig:ds:classify}
\end{figure*}

\pb{IP prefix filtering.} \camfix{(step \circled{3b} in Figure~\ref{fig:ds:filter})}\reviewfix{E3, F1}  The previous analysis indicates that ASN filtering is not enough to identify network measurements associated with SNOs. Out of 41 SNOs, only measurements for OneWeb and O3b can be identified using only their ASNs. While some ASNs can be filtered as clearly departing from the expected latency pattern (\eg 27277 for Starlink and 201554 for SES), other ASNs are harder to filter as they exhibit  mixed latency distributions within the same ASN (\eg 10538 from TelAlaska). As a next step, we dissect the latency characteristics when considering  IPv4 prefixes. In our dataset, there are two main IP blocks as per M-Lab annotation~\cite{mlab-annotation-service}: /24 (66.7\%)  and /21 (21.9\%). Given it is the smallest and most popular IP block, we consider /24 IPv4 prefixes to group latency values. Note that the dataset only contains IPv6 data for four operators: HughesNet, SSI, TelAlaska, and GlobalSat.  With the exception of HughesNet, this data collectively represents less than 0.1\% of the full dataset. In case of HughesNet, we do not find any discernible pattern in terms of prefixes and latency. Therefore, we exclude IPv6 data from this filtering step. %

Since no LEO-based SNO is left, we introduce only MEO (latency > 200ms, or 10th percentile from O3b latency distribution) and GEO (latency > 500ms~\cite{perdices2022satellite}) filters. With this strict filtering, we retain /24 prefixes whose speed tests \textit{only} exhibit latencies within the above filters, and which have at least 10 speed tests. This prefix-based filtering retains less than 1\% of the remaining speed tests, and it spans 25 /24 prefixes from 6 SNOs (Figure~\ref{fig:ds:prefix}). The figure shows five GEO operators and one operator (SES) supporting both MEO and GEO. Note that SES acquired O3b, the only MEO operator in our dataset, in 2016~\cite{ses2023exercise}; in the following, we refer to O3b/SES as a combined MEO operator.

\pb{Relaxing the prefix-based filtering.} \camfix{(step \circled{3c} in Figure~\ref{fig:ds:filter})}\reviewfix{E3, F1}  Figure~\ref{fig:ds:lateny} shows the latency distribution for the remaining IP ranges after prefix filtering for Viasat. The figure shows that 75.105.63.0/24 was discarded just due to few outliers -- thus indicating the prefix filtering might be indeed too strict -- while the other three prefixes have GEO-like median latencies (around 500/600~ms) but also very wide latency  distributions, \eg 45.232.115.0/24 has 30\% of latency values smaller than 70.6ms. The inset of Figure~\ref{fig:ds:lateny} shows the evolution over time of the latency values for one specific IP/user in the prefix 45.232.115.0/24. The figure shows three \textit{clusters} of latency values centered around 600ms (GEO-like), 100-150ms, and 20-40ms. Clearly, such low latencies are incompatible with a GEO-based connection, and likely representative of a terrestrial connection for which the satellite link acts as a backup. This trend matches online evidence of SNOs, especially operating in the GEO orbit, using the satellite link as a \textit{backup} for some unreliable wired access offered by the same operator~\cite{viasatback, hughesback}. 

Motivated by the above observation, we relax the prefix-based filtering, thus tolerating mixed latency patterns within /24 prefixes, or even for a single IP address. For the 6 SNOs covered by the prefix-based filtering (see Figure~\ref{fig:ds:prefix}), we further allow speed tests with latencies bigger than the minimum latency observed, \eg 548.9ms for Viasat (blue-dotted line in Figure~\ref{fig:ds:lateny}). For the other SNOs, we use the minimum latency across the 6 SNOs covered by the prefix-based filtering (527ms). 

\pb{SNO discovery in public traces.} \camfix{(final data accumulation step \circled{4} in Figure~\ref{fig:ds:filter})}\reviewfix{E3, F1} After applying the above strategy on M-Lab traces (see Section~\ref{sec:meth:datasets}), we identify 18 SNOs (2 LEO, 1 MEO, and 15 GEO) with a minimum of 34 speed tests (Kacific) and a maximum of 11.7 million measurements (Starlink), as summarized in Table~\ref{tab:snos} . Instead, for \ripe, we identify 67 probes on Starlink, 6 on Viasat, and 2 on HughesNet. However, Viasat and HughesNet probes were all inactive during the past year or more according to \ripe website~\cite{ripeatlas}. Hence, we restrict the \ripe analysis to Starlink only, whose probes are distributed across 15 countries. Table~\ref{tab:ripe_results} shows the number of \texttt{traceroute} measurements per country, with the US having the highest number of measurements (about 3M), followed by Australia (460k). \camfix{We add that due to the nature of the available data and the diversity of testers enrolled, we could not study all SNOs with equal depth. However, we have made every effort to conduct a thorough analysis with the available data.}\reviewfix{E1, D2}

\begin{table*}[]
\centering
\footnotesize
\setlength{\tabcolsep}{4pt}
\begin{tabular}{|l|l||l|l||l|l||l|l||l|l|}

 \hline
SNO & \# access & SNO & \# access & SNO & \# access & SNO & \# access & SNO & \# access \\
\hline
Starlink & 11.7M &   TelAlaska & 3.05K     & KVH & 951       & Avanti & 122    & Isotropic & 35\\
 O3b/SES & 78.1K   & OneWeb & 2.95K  & SSI & 260       & IntelSat & 91   & Kacific &   34\\ \cline{9-10}
Viasat & 50K   & HughesNet & 2.80K & Eutelsat & 235  & Hellas-Sat & 48 \\
 SES & 23.2K       & Marlink & 1.42K   & GlobalSat & 135 & Ultisat &   37  \\
 \cline{1-8}
\end{tabular}
\caption{Filtered SNOs and the total number of access from each SNO. Starlink \& OneWeb operates in the LEO orbit, O3b/SES in MEO, and the rest operate in the GEO orbit.}
\label{tab:snos}

\end{table*}

\begin{table*}[]
\centering
\footnotesize
\begin{tabular}{|l|c|c|c||l|c|c|c||l|c|c|c|}
\hline
Country & \# Probes & Start & \# trace- &  & \# Probes & Start & \# trace- &  & \# Probes & Start & \# trace- \\ 
 & & time & routes & & & time & routes & & & time & routes \\ \hline
AT & 2  & 22/05 & 0.24M & DE & 5  & 22/05 & 0.71M & NL & 3  & 22/05 & 0.38M \\
AU & 4  & 22/05 & 0.46M & ES & 2  & 22/06 & 0.10M & NZ & 1  & 22/05 & 0.22M  \\
BE & 1  & 23/01 & 0.07M & FR & 5  & 22/11 & 0.35M & PH & 1  & 23/03 & 0.02M \\
CA & 2  & 22/05 & 0.28M & GB & 5  & 22/08 & 0.29M & PL & 1  & 23/01 & 0.06M \\
CL & 1  & 23/02 & 0.05M & IT & 1  & 22/10 & 0.12M & US & 33 & 22/05 & 3.08M 
\\\hline
\end{tabular} %
\caption{Summary of \ripe dataset.}
\label{tab:ripe_results}
\end{table*}

\subsection{Prolific Census}
\label{sec:meth:prolific}
Prolific~\cite{prolific} is an online crowdsourcing platform with more than 130,000 vetted testers. Prolific offers powerful prescreening APIs for recruiting testers based on demographics, location, and even Internet Service Provider (ISP). ISP-based prescreening is only available for testers located in the US and the UK, where Prolific claims to be capable of correctly verifying ISP information~\cite{prolific-blog}.

According to Prolific's prescreening at the time of this study, 160 testers subscribe to any one of three SNOs: Starlink, HughesNet, and Viasat, \ie no other SNO is detected in their list of available operators. We recruited these 160 testers for a survey hosted on our server, where we ask them to provide the name of their provider, their location (city/state), and a score of how satisfied they are with their service. We ran the survey for seven days, attracting 30 participants. We found that only 20 testers were connecting from an IP address belonging to the above SNOs. Intuitively, it can be hard for Prolific to constantly verify how their testers are connected. For example, a tester can create an account at home but then participate in surveys from her phone or from work. This suggests a limitation in Prolific's prescreening, but it also means that potentially more Prolific testers might have access to a satellite-based connection.

Based on the latter observation, we created multiple measurement campaigns over the course of two weeks, where we do not leverage Prolific prescreening, but request, in the study description, that only testers connected via an SNO participate. At the same time, we enable IP address-based access control at our server (via nginx~\cite{nginx}) which allows only IP addresses belonging to SNOs from Table~\ref{tab:snos}. Our studies attracted a total of 14,371 Prolific testers, out of which 57 were actually connected via  Starlink, HughesNet, and Viasat. We find that, overall, Starlink users are much more satisfied with their service than both HughesNet and Viasat users (see Figure~\ref{fig:census:score} in the Appendix). For example, only one user out of 20 considers Starlink connectivity as ``poor'', while most users consider it either good or very good. On the other hand, ``ok'' is the highest score reported for both HughesNet (55\% of the answers) and Viasat (18\% of the answers). 

\subsection{Limitations}
\label{sec:meth:limitations}
\pb{Lack of ground truth.} The accuracy of the methodology described in Section~\ref{sec:meth:identify}
to identify measurements performed via SNO is hard to quantify. This is because we lack  ground truth to compare against, and we thus rely on a comparative analysis between operators latency profiles derived from M-Lab data. This means that while our methodology catches obviously incorrect associations between ASNs and SNOs (see Figure~\ref{fig:kde:starlink} for example), as well as IP reusage across non-satellite access (see  Figure~\ref{fig:ds:lateny}), it can potentially introduce some errors especially when the latency difference between technologies is small such as when comparing LEO and MEO. However, our analysis in Section~\ref{sec:meth:identify} shows that LEO operators are easy to identify, and most of the remaining operators are GEO-based, whose latency profile is hard to confuse with LEO or terrestrial accesses due to the long latency inherent to GEO-based connectivity. We thus believe that our methodology catches most major errors, but we acknowledge that further unknown errors might exist. \camfix{We opt not to perform any geographical analysis of end users  using the M-Lab data due to the known fallacies of geo-location mapping~\cite{poese2011ip, gouel2021ip, callejo2023deep}. Further, this introduces an additional layer of complexity in our data handling process. However, we perform spatial dissection for data collected from RIPE probes and Prolific users, as their locations are known.}\reviewfix{B2}

\pb{Speedtest accuracy.} NDT measurements on M-Lab rely on a single TCP connection, and it is well-known~\cite{feamster2020speedtest} that single-flow TCP often under-estimates the available bandwidth due to slow-start. Further, owing to frequent satellite handoffs, it is possible for LEO satellite communication networks to have temporary periods of high packet loss which are not related to congestion~\cite{kassem2022browser}. If such losses occurred during the NDT tests in our dataset, it would have led to the TCP flow decreasing its sending rate and consequently underestimating the bandwidth. This is a known artefact of TCP-based speed-tests, but from our dataset it is not possible to make an ex post facto analysis to filter out tests which suffered losses due satellite handoffs etc. In addition we focus exclusively on download tests for two reasons. Firstly, the kernel level \texttt{TCP\_Info} is solely captured at the server, ensuring the reliability of the values for download measurements. Secondly, the upload and download traces are obtained separately, necessitating the use of a time-interval based heuristic to merge the upload and download tests from the same client~\cite{paul2022importance,sundaresan2017challenges}. As also noted in~\cite{jain2022ukranian}, we acknowledge that users typically run speed tests when they experience degraded network conditions. It follows that  results derived from M-Lab speed tests might underestimate the actual network conditions. \camfix{We want to highlight that the presence of middleboxes, VPNs, NAT, or any other end-user configurations (such as speed throttling at the browser) could potentially affect the results we obtain. However, we argue that such variations are unavoidable in real-world data traces, and these data anomalies should be considered as outliers that may not necessarily have a significant impact on the overall trends.}\reviewfix{A2,B1,D1}\looseness=-1 %

\section{A Bird's-Eye Look Into SNOs}
\label{sec:bird}

\begin{figure*}

 \subfloat[\texttt{RTT across time and SNOs.\reviewfix{E4}}]{\includegraphics[width=0.32  \linewidth]{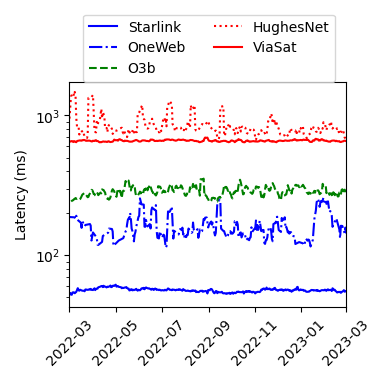}
\label{fig:sno:rtt:time}}
\subfloat[\texttt{Normalised and actual (inset) variation in jitter.}]{
\includegraphics[width=0.32 \linewidth]{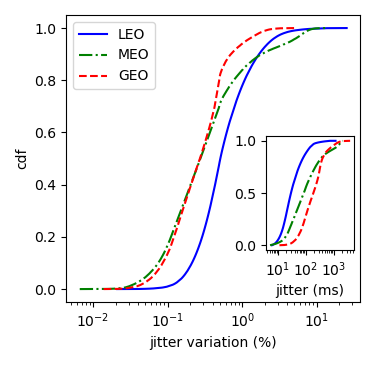}\label{fig:sno:jitter}}
\subfloat[\texttt{\% bytes retransmitted.}]{
\includegraphics[width=0.32 \linewidth]{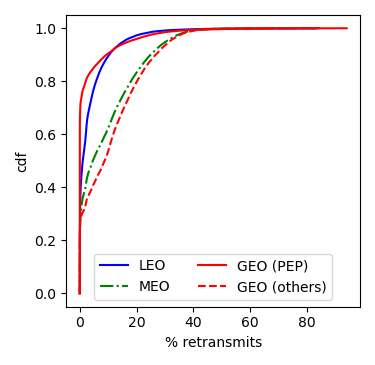}\label{fig:sno:retx}}
\vspace{-0.1in}
\caption{Performance of LEO, MEO and GEO SNOs.}

\end{figure*}

\pb{Latency distributions.} As shown in Figure~\ref{fig:KDE}, SNOs' performance is expected to vary depending on the orbit they operate in. Figure~\ref{fig:sno:rtt} shows boxplots of the (access) latency per SNO  using the M-Lab dataset.  As expected, the latencies for GEO SNOs are higher, with a median of 673.5 ms, followed by MEO (279.36 ms)  and LEO (56--154ms). Among the two LEO operators, Starlink is overall 3x faster than OneWeb. When focusing on GEO operators, the figure shows a significant difference between the best performing GEO SNO (SSI, with a median latency of 620.4 ms) versus the worst performing one (KVH, with a median latency of 835.2 ms).

Next, we analyze the evolution over time of the access latency for the 5 most popular SNOs per orbit: Starlink, OneWeb, O3B, HughesNet, and Viasat. Figure~\ref{fig:sno:rtt:time} shows the median  access latency per day and operator over the last year. The figure confirms the trend observed in Figure~\ref{fig:sno:rtt}, at any point in time. It further shows that Starlink and Viasat, \ie the two most prominent operators in their respective orbits, are quite stable, with daily latency variation (95$^{th}$ \%ile) up to 3.1\% and 7.2\%, respectively. O3b shows a similar trend, although its daily variation increases to 41.4\%. Both HughesNet and OneWeb are instead often affected by significant daily latency variations, up to  72\% for HughesNet and 120\% for OneWeb.

\pb{Variability in latency.} Next, we investigate the impact of variability in latency (\textit{jitter}) when accessing SNOs operating in different orbits. In order to take into account the large differences in the overall latencies between LEO, MEO and GEO, we normalize the jitter ($95^{th}$ percentile) relative to the $5^{th}$ percentile latency for each M-Lab request, showing how the jitter varies as a fraction of the lowest latencies achievable in each orbit. Figure~\ref{fig:sno:jitter} shows Cumulative Distribution Functions (CDFs) of this \textit{jitter variation} per orbit. The figure shows that LEO exhibits a greater degree of jitter variability compared to GEO, with a median of 0.5 as opposed to 0.28. This observation indicates that, even though GEO SNOs are characterized by overall much higher latencies than LEO SNOs, they are overall more stable. This discrepancy can be attributed to the fact that LEO satellites require   frequent satellite handoffs~\cite{park2021trends, akyildiz1999handover}. MEO is characterized by less frequent satellite handoffs than LEO~\cite{park2021trends} and shows jitter variability similar to GEO. However, the figure also shows that, for 10\% of the values, MEO's jitter variation is comparable to and even higher than LEO. This suggests that, when there is a handoff, recovering from it is more difficult in MEO networks; this happens because there are fewer satellites deployed than in the LEO satellite mega constellations. \camfix{Note that, while this demonstrates the network's variability in connection based on latency, we observe that LEO outperforms GEO when comparing absolute jitter values, as shown in Figure~\ref{fig:sno:jitter} (inset) --- over 80\% of the GEO trace exhibited a jitter of 100ms or more, while LEO had less than 20\% with such high jitter.} \reviewfix{E5, F3}

\pb{Retransmissions.} Finally, we  shift our attention to a major side-effect of high latency and variability in latency: packet retransmissions. Figure~\ref{fig:sno:retx} shows the percentage of bytes that required retransmission. LEO satellites have fewer retransmissions than MEO, which is expected considering that the lower latencies to LEO orbits can lead to fewer retransmission timeouts and errors. The picture for GEO operators is more nuanced. One class of operators, shown as ``GEO (others)'' have a significant (median 8.74\%) fraction of data retransmitted, which can be expected in GEO, given the difficulties of transport protocols on high latency asymmetric links. However, four GEO operators (HughesNet, Viasat, Eutelsat and Avanti) are exceptions to this trend and their retransmission fractions are close to those of LEO operators. Their retransmission factor CDF is shown as the line ``GEO (PEP)'' in Figure~\ref{fig:sno:retx}, as we notice that all four operators use Performance Enhancing Proxies (PEPs)~\cite{rfc2488, rfc3449}\footnote{HughesNet~\cite{hughes-pep}, Viasat~\cite{viasat-sna}, Eutelsat~\cite{kuhn2020quic} and Avanti~\cite{netsys19}.}. PEPs manage the TCP connection on either side of the high latency bent-pipe link and thereby mitigate the effect of high latencies leading to retransmission timeouts. \camfix{If there are any latency-related retransmissions on the satellite link, these can be  managed by the PEPs so that they are transparent to the TCP connection between the end hosts being connected~\cite{rfc3135}.}\reviewfix{F5} Figure~\ref{fig:sno:retx} points to the effectiveness of this strategy.

\pb{Geographic connectivity characterization.}
\label{bgp_analysis}
End-to-end latency in an SNO can be influenced not just by the orbit it operates in, but also how well its network infrastructure on the ground is provisioned, \eg the number/location of its PoPs and their physical connectivity to the SNO's upstream provider networks. However, mapping the geographic coverage of an SNO can be challenging. For the majority of SNOs, there is no publicly available infrastructure map. \camfix{As of this writing,} we only find PoP location information for Starlink~\cite{starlinkmap}, SES~\cite{sesmap} and Hellas-Sat~\cite{hellas-sat}. In addition, from reverse DNS lookup of 900K SNO IP addresses discovered from M-Lab datasets, we find that no SNO other than Starlink encodes PoP location information in subscriber domain names. Finally, the client location information available in M-Lab datasets may not necessarily represent the PoP locations of SNOs, but simply the (approximate) locations of clients inferred by Google's location service. \looseness=-1

Given these challenges, our approach is to ``indirectly'' infer the geographic coverage of SNOs from BGP peering data. Our intuition is the following. Since none of the existing SNOs is a tier-1 ISP, they must peer with  larger upstream ISPs to obtain global reachability of their ground infrastructure.  If an SNO peers with upstream ISPs in many different locations, that should be an indication of similar geographic spread of its ground stations.  Based on this intuition, we analyze BGP peerings of SNOs with the BGP route-views~\cite{routeviews} collected from similar periods as our SNO datasets. 

Initially, we tried to identify exact physical locations of SNO's BGP peering points at available Internet exchanges~\cite{peeringdb}.  However, we find that SNO information at public exchange points is not very complete except for very few SNOs with many BGP peers, such as Starlink and SES.  For broader coverage, we end up using peering ASes' country jurisdiction as the approximate peering location. AS country information is from the Regional Internet Registries, where the ISO 3166 country codes are added during ASN assignment~\cite{autnums}. When we cross-check the inferred country-level PoP locations with available ground-truth PoP location data of several SNOs, we find that our approach discovers \camfix{10 out of 30}, 7 out of 22, and 2 out of 2 countries for Starlink, SES and Hellas-Sat, respectively. The discovered countries cover \camfix{74\%}, 57\% and 100\% of their city-level PoP locations, respectively. \camfix{In case of Starlink, missing country-level PoP locations include European countries (\eg Spain, Portugal, France, Norway), South American countries (\eg Mexico, Brazil, Chile, Peru) and Caribbean countries (\eg Fiji, Dominican Republic).  Upon closer inspection of the discrepancies, we find that some of Starlink's peering ASes actually have \emph{continent-wide} presence in their network connectivity (\eg AS1299 (Arelion) and AS6762 (Telecom Italia Sparkle) for Europe and AS7195 (EdgeUno) for South America).  Hence, Starlink may peer with these neighbors in multiple country locations within the continents, beyond the single jurisdiction registered with these AS numbers. In fact, multi-location peerings are reported in CAIDA's peering geolocation data set~\cite{caidageo}, where we observe that about 6\% of peering links are associated with more than one location.  Unfortunately, we could not use the CAIDA dataset because its collection period (2016) is considerably earlier than ours (2021--2023), and quite a few new ASes have appeared since then.}\reviewfix{A3, E6, F6} Although our approach can underestimate the geographic coverage of SNOs due to this limitation, its main benefit is that it enables comparative analyses of geographic coverage of \emph{all existing SNOs}, as well as longitudinal studies on the \emph{historical evolution} of a specific SNO's ground infrastructure.

\begin{figure*}[tb]
    \includegraphics[width=0.9\linewidth, trim=0mm 220mm 0mm 5mm, clip=true]{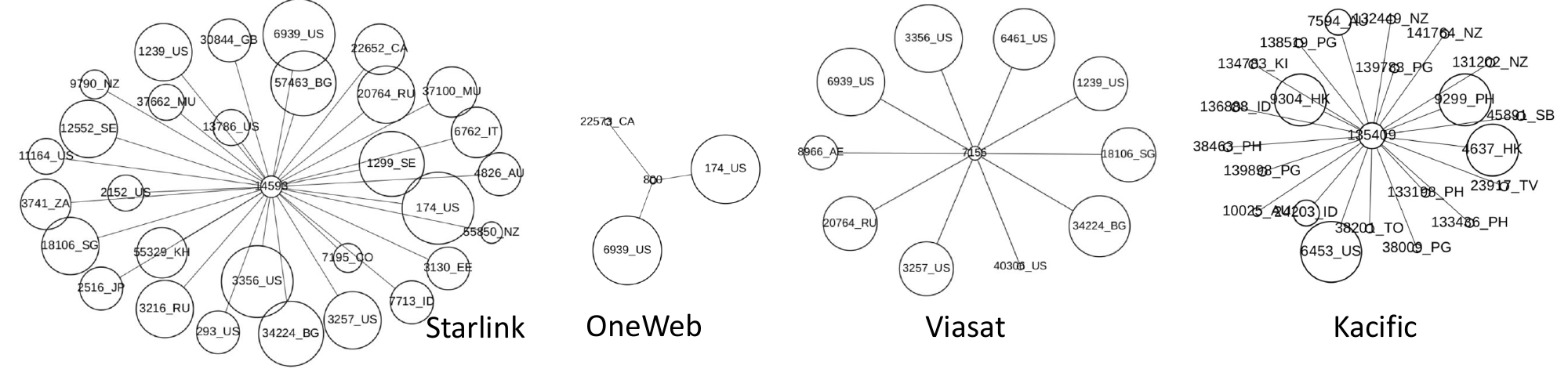}
    \caption{BGP peering visualization of SNOs (based on BGP route-views collected on 2023/1/1).}
\label{fig:bgp_viz_small}
\end{figure*}

Figure~\ref{fig:bgp_viz_small} visualizes BGP peerings of few SNOs; more examples can be found in Figure~\ref{fig:bgp_viz} in the Appendix. For a given SNO, the label of each peering AS indicates its ASN and country. The diameter of each peering AS indicates it ``size'', estimated  with its node degree. From relative AS sizes, we can speculate whether or not a peering neighbor is an SNO's upstream provider~\cite{gao01}. For example, Starlink is connected to AS3356 (Level3) which is much bigger than Starlink, so this AS is likely a Starlink's upstream provider. Conversely, Kacific peers with much smaller ASes than itself, which we confirm are small regional ISPs obtaining satellite access from Kacific~\cite{kacific}.

From geographic characterization of SNO PoPs, we make the following observations.
First of all, the upstream connectivity of SNOs in both LEO and GEO categories is significantly varied.  For example, in GEO category, Hellas-Sat and UltiSat are not connected to any tier-1 providers, while Viasat and Kacific are well-connected to multiple tier-1 providers.
Similarly, in the LEO category, Figure~\ref{fig:bgp_viz_small} shows that Starlink is peering with major upstream providers all around the globe, while OneWeb is connected to only two US-based providers. This could potentially explain the  significant latency difference between them in Figure~\ref{fig:sno:rtt}. We also observe (not shown) that Starlink shows consistent latency performance worldwide, while OneWeb exhibits skewed performance in North America vs.~the rest of the world. The SNO in MEO category (\eg SES) tends to be much better connected to major transit providers than pure GEO-based SNOs. MEO-based SNOs are likely to pursue more aggressive peering to cover more grounds as they expand to LEO satellites. Finally, we observe that different SNOs exhibit widely varying growth trajectories in their PoP infrastructures over time, \eg Starlink had explosive growth compared to HughesNet which was stagnant between 2021 and 2023 (Figure~\ref{fig:bgphistory}); such geographic diversity may help Starlink as it expands to new countries~\cite{reuters}. We explore Starlink latencies in different countries next.

\section{A Closer Look at Starlink}
This section zooms in on Starlink using built-in measurements performed by 67 Ripe Atlas probes between May 2022 and May 2023 (see Table~\ref{tab:ripe_results}). Our goal is to shed some light on Starlink infrastructure worldwide and its impact on network latency. %

\begin{figure*}[tb]
    \subfloat[\texttt{RTT to Starlink PoPs.}]{\includegraphics[width=0.33\linewidth]{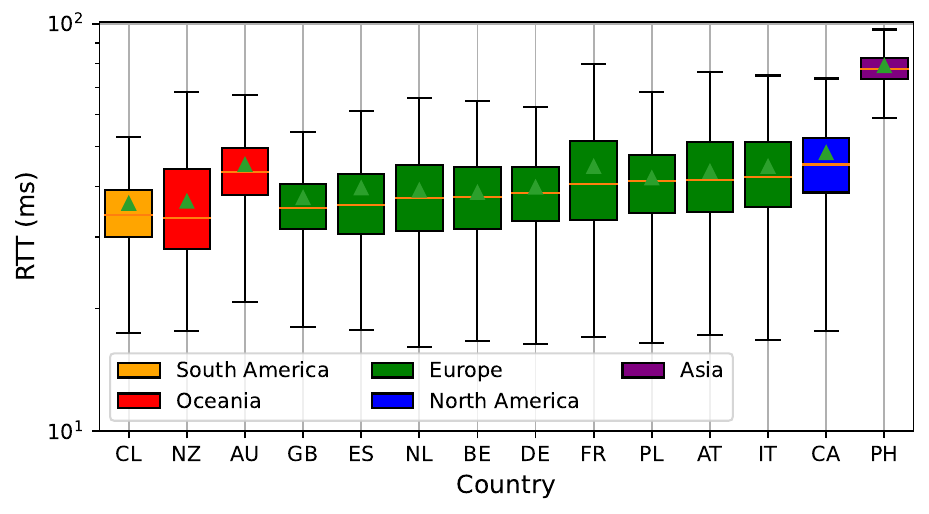}\label{fig:ripe_global_first_hop_RTT}}\hfill
    \subfloat[\texttt{RTT to root DNS servers.}]{\includegraphics[width=0.33\linewidth]{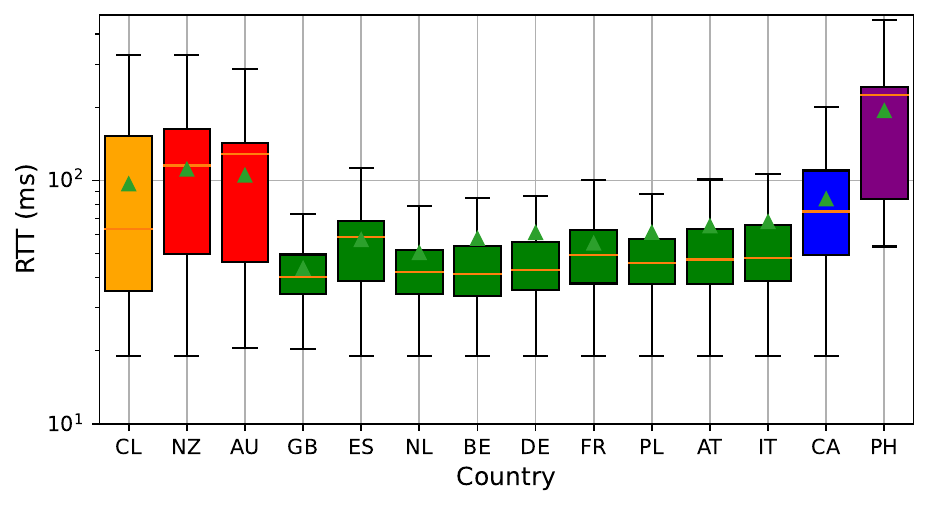}\label{fig:ripe_last_hop_RTT}}\hfill
    \subfloat[\texttt{Hops to root DNS servers.}]{\includegraphics[width=0.33\linewidth]{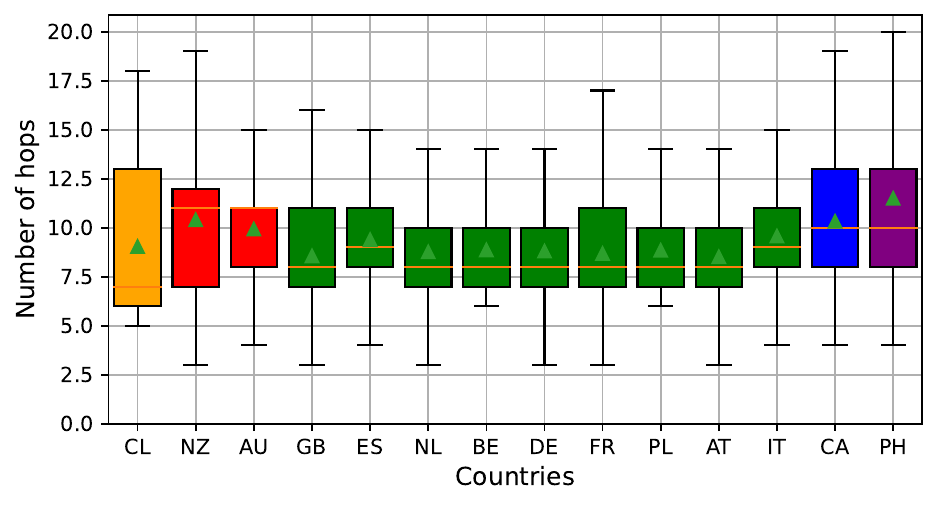}\label{fig:ripe_num_hops}}\hfill  
    \vspace{-0.1in}
    \caption{Analysis of \texttt{traceroute} measurements between Ripe Atlas probes on Starlink and root DNS servers; Rest of the world (34 probes, in total).}
\label{fig:ripe_cdfs}
\end{figure*}

\pb{Rest of the world.\footnote{``Rest of the world'' refers to worldwide, excluding the United States.}} Figure~\ref{fig:ripe_global_first_hop_RTT} shows boxplots of the RTT measured over one year between \ripe probes located outside the US (34 probes) and Starlink PoPs; see Table~\ref{tab:ripe_results} for the number of measurements per country. Results are organized by country and grouped by continent using the following color scheme: green for Europe, blue for North America (Canada only), yellow for South America, purple for Asia, and red for Oceania. The figure shows that New Zealand and Chile probes exhibit the lowest RTT to their respective Starlink PoPs, with median values at approximately 33~ms. Most European countries follow closely with median values of 35-40~ms. Canada and Australia experience slightly higher RTTs, averaging at around 45~ms. The highest RTT to a Starlink ground station is in the Philippines, with a median of 80~ms, almost twice what measured in other locations. 

\begin{figure*}[tb]
    \subfloat[\texttt{US}]{\includegraphics[width=0.33\linewidth]{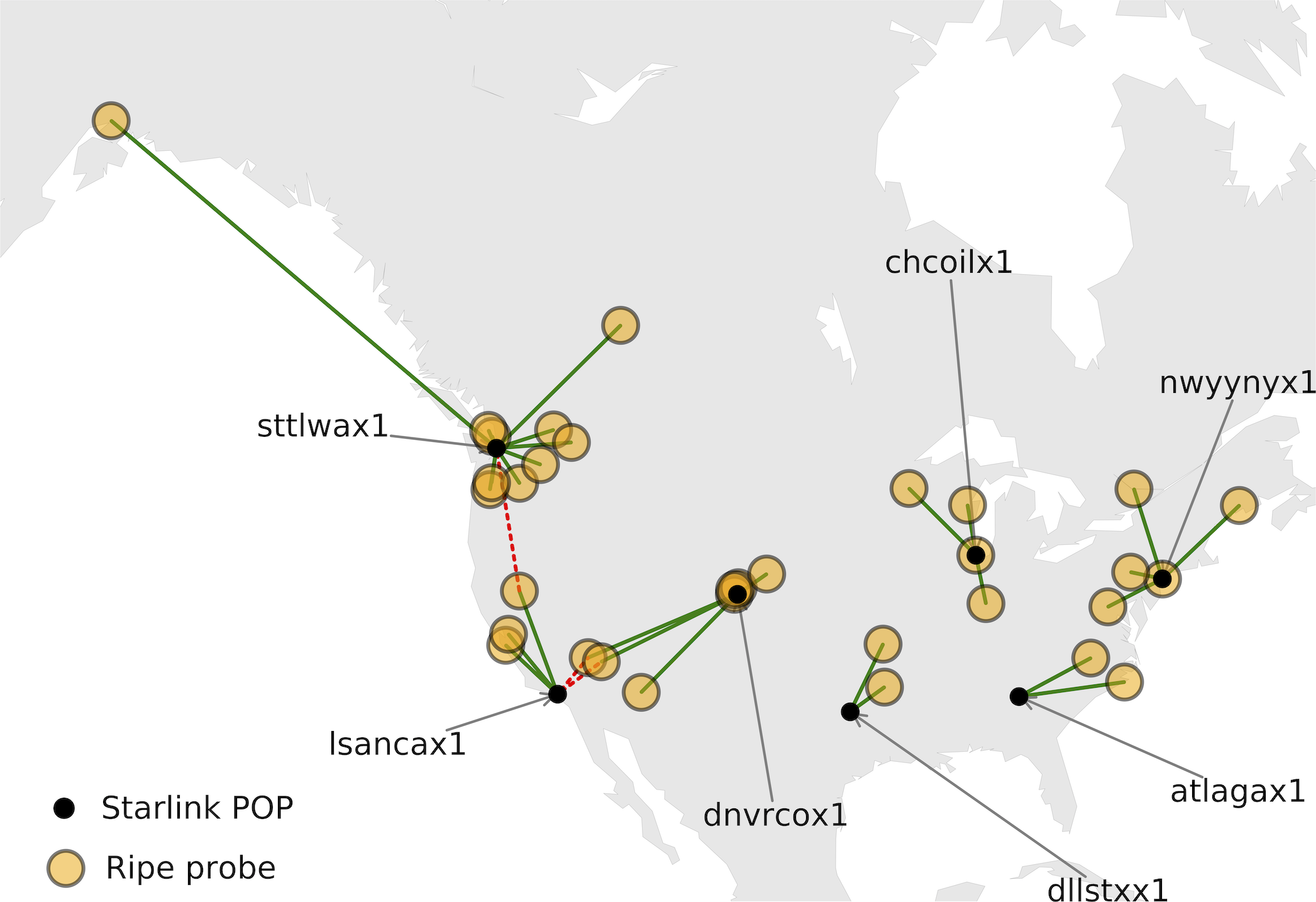}\label{fig:ripe_us_map}}\hfill
    \subfloat[\texttt{Europe}]{\includegraphics[width=0.22\linewidth]{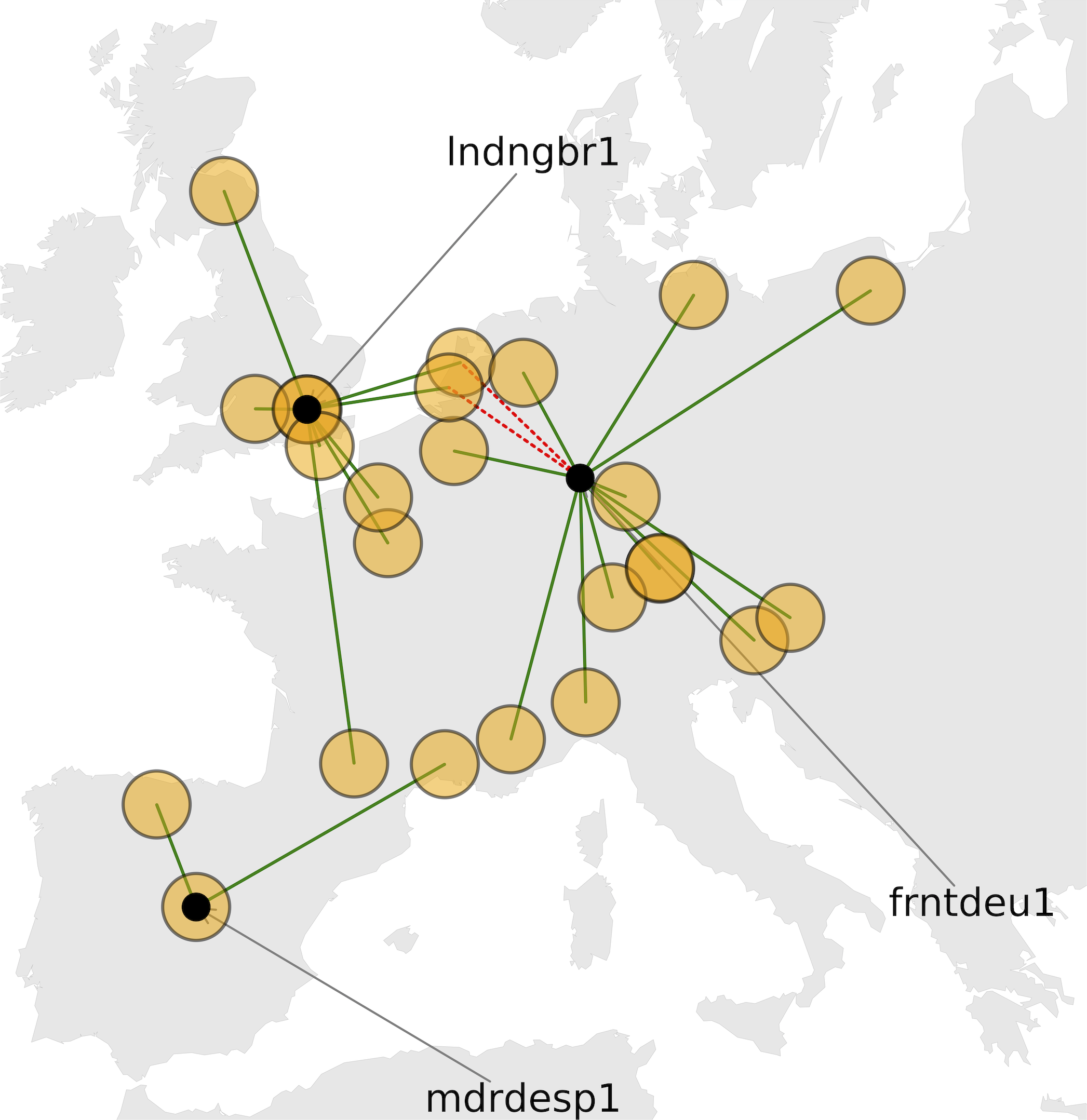}\label{fig:ripe_eu_map}}\hfill
    \subfloat[\texttt{Oceania}]{\includegraphics[width=0.26\linewidth]{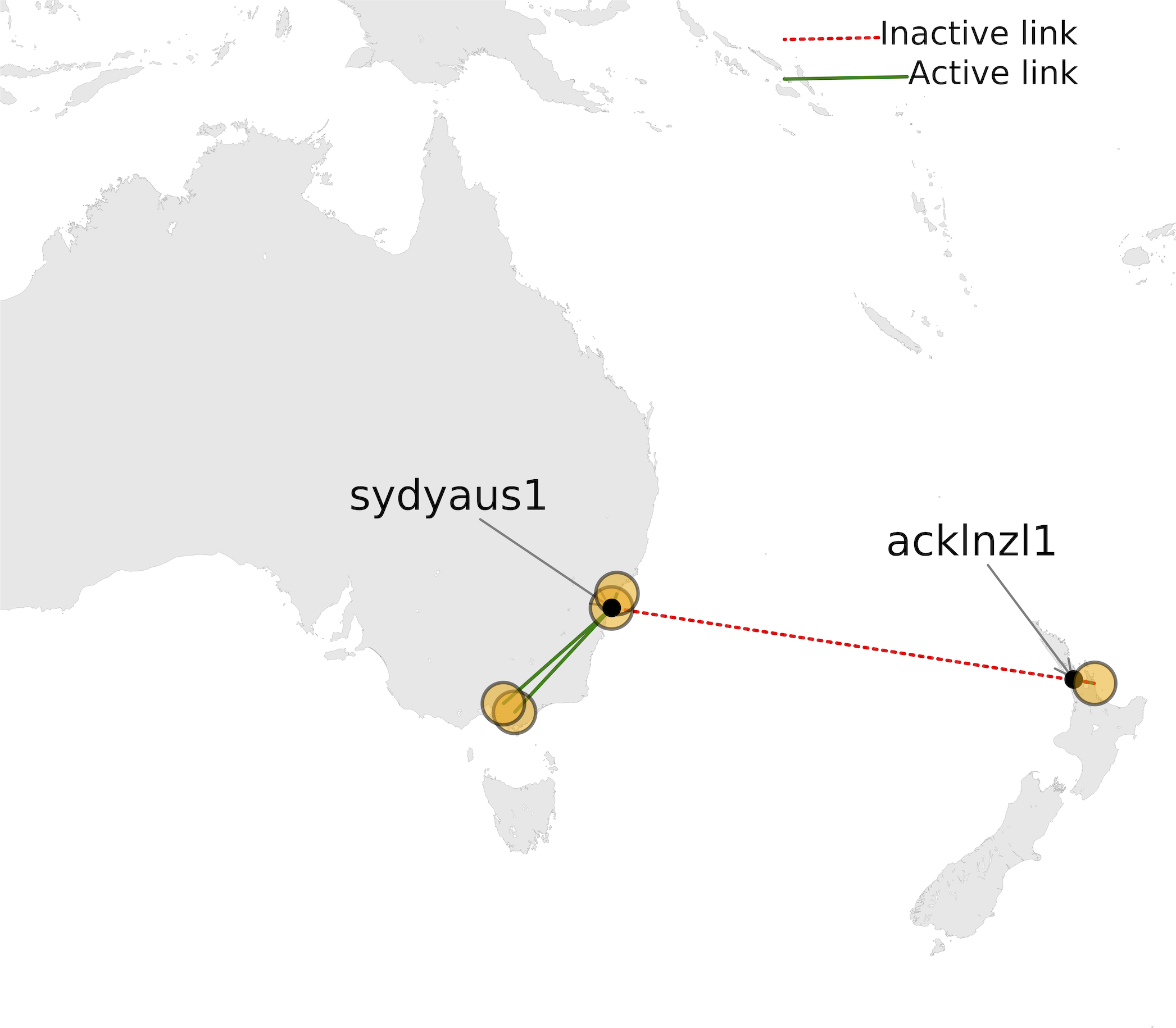}\label{fig:ripe_oceania_map}}\hfill\subfloat[\texttt{Philippines}]{\includegraphics[trim={0 1.6cm 0 0},clip,width=0.16\linewidth]{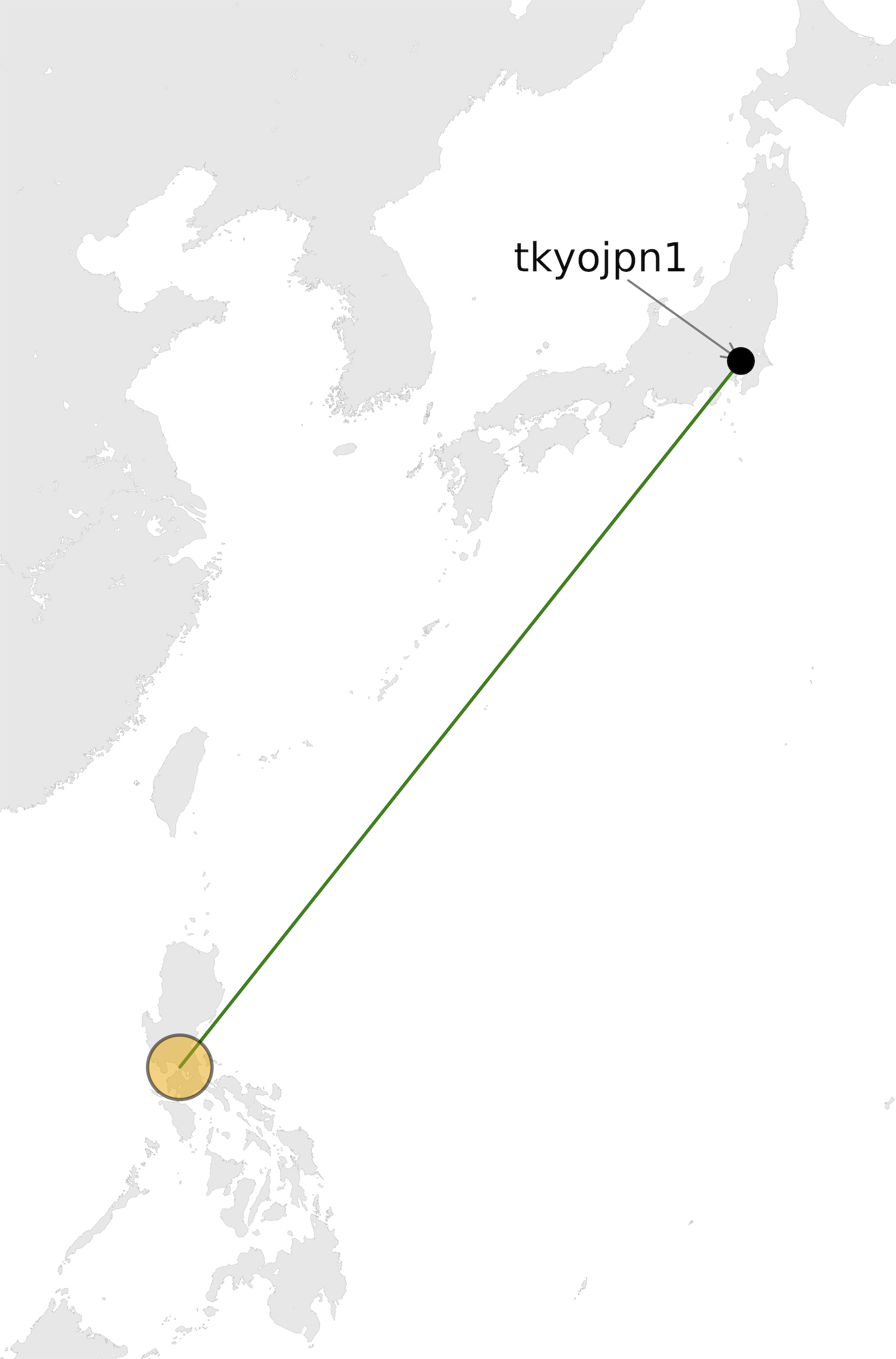}\label{fig:ripe_philippines_map}}\hfill    
    \vspace{-0.1in}
    \caption{Location of \ripe probes and connectivity to their Starlink PoPs. Green line refers to active links as of April 2023. Red dotted lines refer to previous links observed between April 2022 and April 2023.}
    \vspace{-0.22in}
\label{fig:ripe_maps}
\end{figure*}

To investigate the cause behind this substantial RTT disparity, we analyze the geographical placement of Starlink PoPs used by each probe. Figure~\ref{fig:ripe_maps} visualizes the location of \ripe probes (yellow circles) along with the PoPs (black circles) they connect to distinguishing between the US, Europe, Oceania, and Philippines. The green lines show active connections between a probe and a PoP, whereas, the red lines show inactive (\ie previously used but not at the time of writing) connection between a probe and a PoP. We do not show the probe-PoP pair in Chile due to space limitation and since they are co-located (approximately 75~km away), thus explaining the very low latency. When focusing on the Philippines probe, located in Manila, the figure shows that the PoP for this probe is located in Tokyo, Japan (``customer.tkyojpn1.pop.starlinkisp.net'') likely causing the extra delay (roughly 40~ms). While we cannot measure the link between Starlink antennas -- located in the Philippines according to Google Maps -- and the Tokyo PoP, we checked the RTT from Manila to Tokyo~\cite{wonderTokyo} and found fairly high values, 177~ms on average.

We now move to the the analysis of RTT to root DNS servers (see Figure~\ref{fig:ripe_last_hop_RTT}). The figure shows negligible difference between European countries, with median RTT values comprised between 40 and 49~ms (apart from Spain that has a slightly higher median RTT of 58~ms); this is expected given their comparable RTT to their respective Starlink PoPs and the presence of many root DNS servers in Europe. Despite having the lowest RTT to a Starlink PoP, the Chilean node experiences an extra 10-20~ms delay to root DNS servers making its performance comparable to the Canadian probe. This happens because only 7 out of 13 root DNS servers are present in Chile, requiring longest routes for about 50\% of the requests, which also explain the wide distribution of RTT values. This behavior is verified by Figure~\ref{fig:ripe_num_hops}, which shows a wide variability of path to root DNS servers comprised between a minimum of 5 hops (\eg to the L root accessed in Santiago, Chile) and 20 hops or above (\eg to the M root which is currently not present in South America). A similar behavior applies to New Zealand and Australia, which only host few local root DNS servers and, for most queries, requires 100-150~ms and more than 10 routing hops. Philippines trails with the highest delay to root DNS servers (about 200~ms).

\begin{figure*}[tb]
    \subfloat[]{\includegraphics[width=0.5\linewidth]{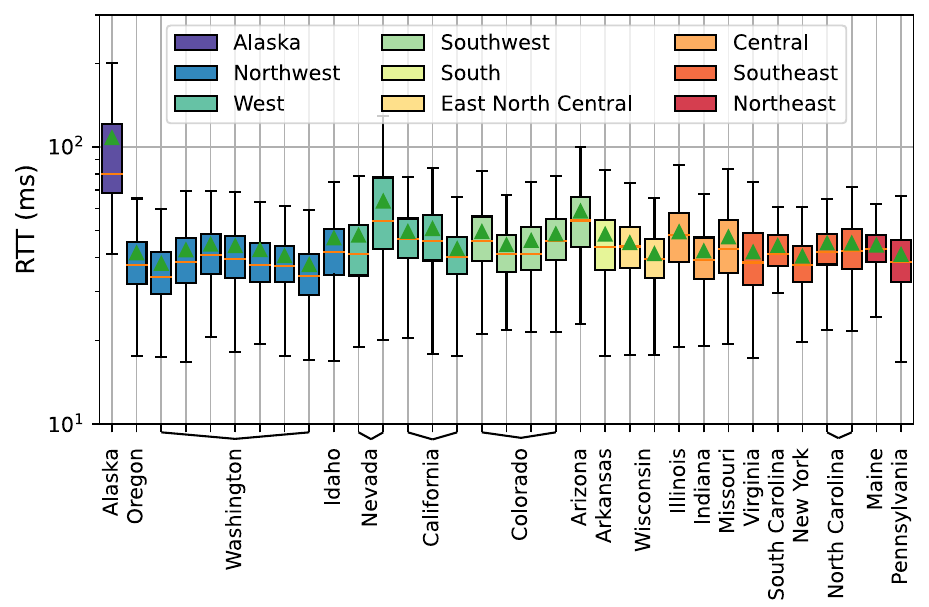}\label{fig:rip_states_cdf_groundstation}}\hfill
    \subfloat[]{\includegraphics[width=0.5\linewidth]{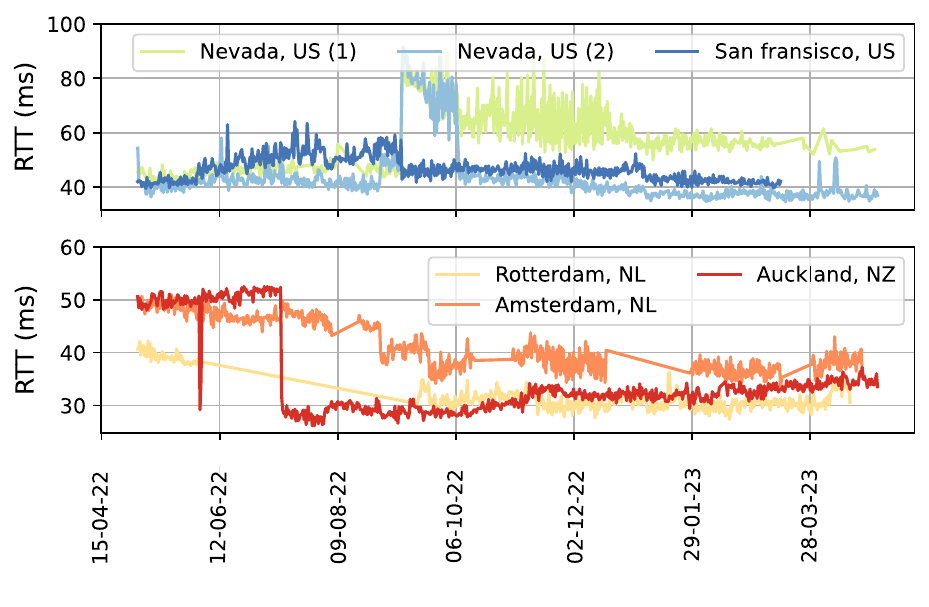}\label{fig:rtt_timeline}}\hfill 
    \vspace{-0.15in}
    \caption{(a) RTT between the 33 \ripe probes located in the US and Starlink PoPs; and (b) RTT over time between (New Zealand, Netherlands, San Francisco, and Nevada) \ripe probes and Starlink PoPs.}
    \vspace{-0.2in}
\end{figure*}

\pb{United States.} Figure~\ref{fig:rip_states_cdf_groundstation} shows boxplots of the RTT measured over one year between \ripe probes located in the US (33 probes) and Starlink PoPs. We only focus on this metric since it is the most relevant to this analysis. Results are organized by state and grouped as follows: Northeast, Southeast, Central, East North Central, South, Southwest, West, Northwest, and Alaska. If we focus on the median, the figure shows that the minimum RTT to  Starlink PoPs sits around 45~ms for the following states: Oregon, Washington, Virginia, New York, and Pennsylvania. Note that this is about 10~ms higher than what observed for the best performing countries in the rest of the world (see Figure~\ref{fig:ripe_global_first_hop_RTT}). The remaining states, with the exception of Alaska, show slightly higher median RTT values, with Arizona reaching a maximum of 55~ms. 

Alaska exhibits a significantly higher RTT than all the other states, with median RTT of 80~ms (75th percentile RTT of 120ms). To understand the underlying cause of this high RTT, we again resort to the analysis of the geographical location of the probe-to-PoP mapping. Figure~\ref{fig:ripe_us_map} shows that the Alaska probe is connected to a PoP in Seattle (Washington State), approximately 2,697 km away, which is likely responsible of the additional RTT incurred by the probe. %

Finally, we analyze Starlink latencies over \textit{time}. For most probes, we observe non statistically significant variations of RTT to their PoPs over one year. Figure~\ref{fig:rtt_timeline} visualizes six probes which instead see a significant change in their Starlink latencies over time. The figure shows that the New Zealand probe experienced a 20~ms latency reduction since July 12th, 2022 when its PoP changed from Sydney (Australia) to Auckland (New Zealand), as indicated by the red dotted line in Figure~\ref{fig:ripe_oceania_map}. Indeed, such PoP change from Sydney to Auckland was tested between June 2nd and 3rd (2022), which is reflected in the sudden latency reduction around the beginning of June. A similar observation can be made for the Netherlands probe, where a 10~ms latency drop was due to a shift from a PoP in Frankfurt to one in London, as shown in Figure~\ref{fig:ripe_eu_map}. Finally, the figure shows also the potential ``damage'' associated with a PoP change. In the case of Nevada, in September 2022 the PoP was changed from Los Angeles to Denver (see Figure~\ref{fig:ripe_us_map}), which caused a 2x increase of the RTT to the assigned PoP. The change was reverted about one month later, allowing the RTT to return to its previous values for one of the two probes in Nevada. No PoP change was detected for the other probe. However, the figure shows a high variable, although down trending, latency for this probe which eventually stabilize around 55ms.

\section{Applications Performance}
Finally, we expand on how \textit{real users} perceive the performance they obtain from SNOs by conducting a user study on real customers of Starlink, Viasat and HughesNet.%
\subsection{Methodology and Data Collection}
\label{app:meth}

\pb{Browser addon design.} As in~\cite{kassem2022browser,varvello2022performance}, we rely on a Chromium addon to perform Internet measurements from end users in the wild. This design choice is motivated by its ease of installation and wide compatibility, as Chromium-based browsers account for 80\% of the market~\cite{chromeMarket}. %
The addon guides the tester through a set of experiments using local ``hooks'' implemented in the addon, providing instructions and requesting the launch of the next measurement as the tester is ready. After each experiment, the addon notifies the tester to close the tab associated with an experiment, if any, and return to the addon. The addon also communicates with our server (the  ``experiment manager'') to be instructed on what experiments should be run, and to report the data collected. Upon completion of a study, the tester is presented with a code which is synchronized (using Prolific's API~\cite{prolificAPI}) with the respective job posted on Prolific. 

We have developed code hooks in the addon for four measurements: Content Delivery Network (CDN) measurements, speedtest (using \textit{fast.com}), webpage loads, and video streaming (using YouTube). %
The speedtest %
experiment relies on the code provided by~\cite{varvello2022performance}, but with slight adjustments to %
enable the upload bandwidth measurement at \textit{fast.com}. When needed, JavaScript (JS) code is injected into a new tab to collect relevant statistics, \eg performance metrics when loading a webpage, and to help a tester, \eg inform when a measurement is completed. %
This injected JS code also communicates with a background script which has permission to: 1) communicate with the experiment manager, 2) collect screenshots (when needed).

After installation, the addon informs the experiment manager of a new tester. At this time, the tester is asked to wait for one minute during a \textit{warm-up} while the background script measures the ``clock drift'' of the client's local clock (along with RTT to our server), and identifies the installed DNS resolver using \textit{https://test.nextdns.io/}. Next, the addon measures the performance of several CDNs (Cloudflare, Google, jsDelivr, StackPath, and Fastly) by fetching a popular JS file (\texttt{jquery.js}) while measuring download time, and collecting HTTP headers. Note that jsDelivr advertises ``optimal'' speed by matching each request to an optimal CDN based on uptime and performance~\cite{jsdelivr}. Finally, the same CDN measurement -- but fetching the minified version of this script (\texttt{jquery.min.js}) -- is repeated meanwhile the tester is asked to provide his/her current location (city and state).

To avoid interference with the browser cache, we request older versions of each file (regular and minified), ranging from \texttt{3.0.0} up to \texttt{3.6.4}, verifying that the file is not served from the cache; if it is, we attempt a different version in this range until no cache entry is found. This allows to perform multiple experiments with the same  tester, and to handle the unlucky event of this file being  previously stored in the cache. We further discard the first file download to realize a DNS primer. The maximum file size variation observed across versions is about 3~KB. When comparing CDN providers, Cloudflare serves the most compressed versions of both files, \ie 28~KB (versus 31-33~KB) for the minified version, and 71~KB (versus 86-89~KB) for the regular version. 

After the initial warm-up, when the tester is ready they can click on a button in the addon which will open a tab pointing to \textit{fast.com}. We use the injected JS code to monitor the statistics on screen, such as bandwidth, latency, and client/server location, while also detecting when the measurement is completed. The next experiment involves loading multiple webpages sequentially, based on a list provided by the experiment manager. The addon automatically iterates through each URL waiting for the \textit{onload} event (or a  \camfix{timeout of about 60 seconds}\reviewfix{E7}) while collecting and reporting performance metrics.  \camfix{As shown in Figure~\ref{fig:prolific:web:TFB}, this timeout was only triggered by one tester from HughesNet (PLT of 62.6 seconds).}\reviewfix{E7} One single screenshot is taken before leaving the tab, which helps verify the data collected. 

\begin{figure*}[tb]
    \subfloat[\texttt{Download speed.}]{\includegraphics[width=0.32\linewidth]{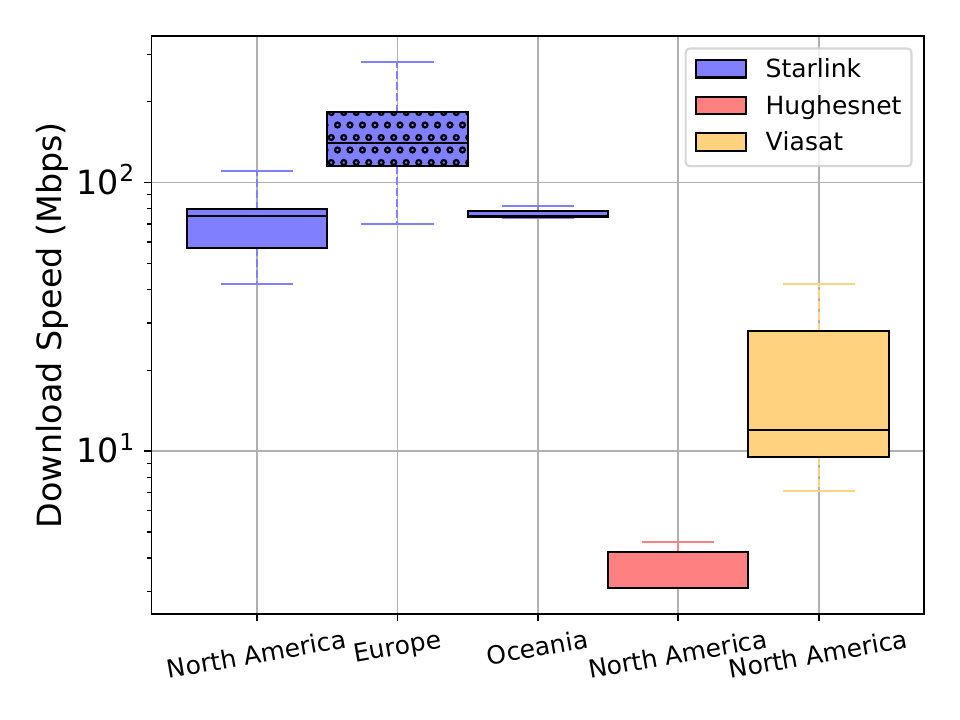}\label{fig:prolific:speedtest:down}}\hfill
    \subfloat[\texttt{Upload speed.}]{\includegraphics[width=0.32\linewidth]{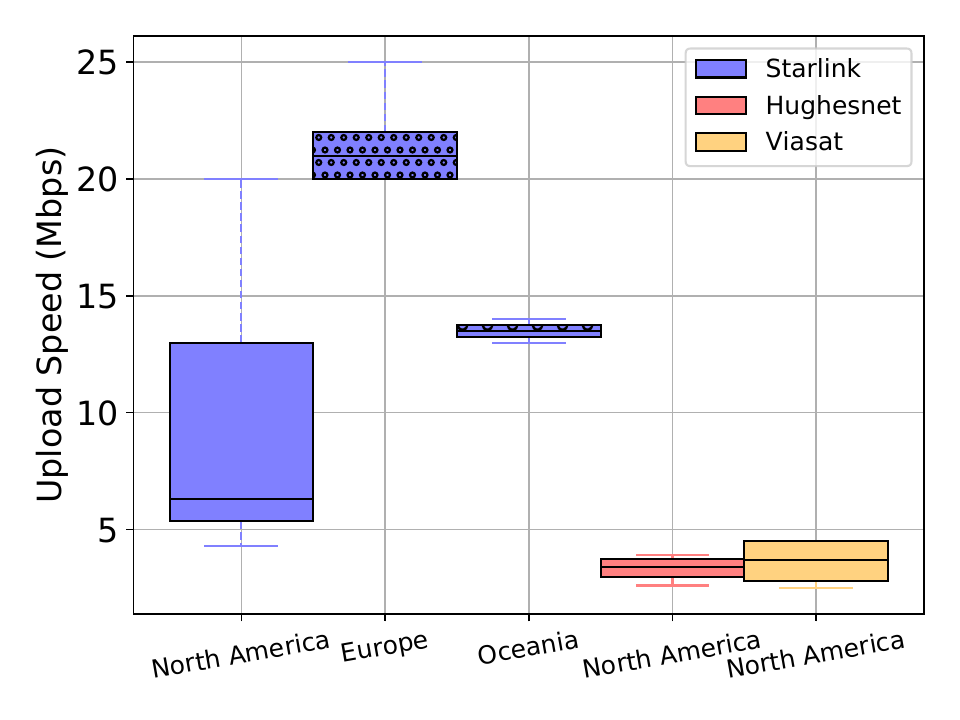}\label{fig:prolific::speedtest:up}}\hfill
    \subfloat[\texttt{Latency.}]{\includegraphics[width=0.32\linewidth]{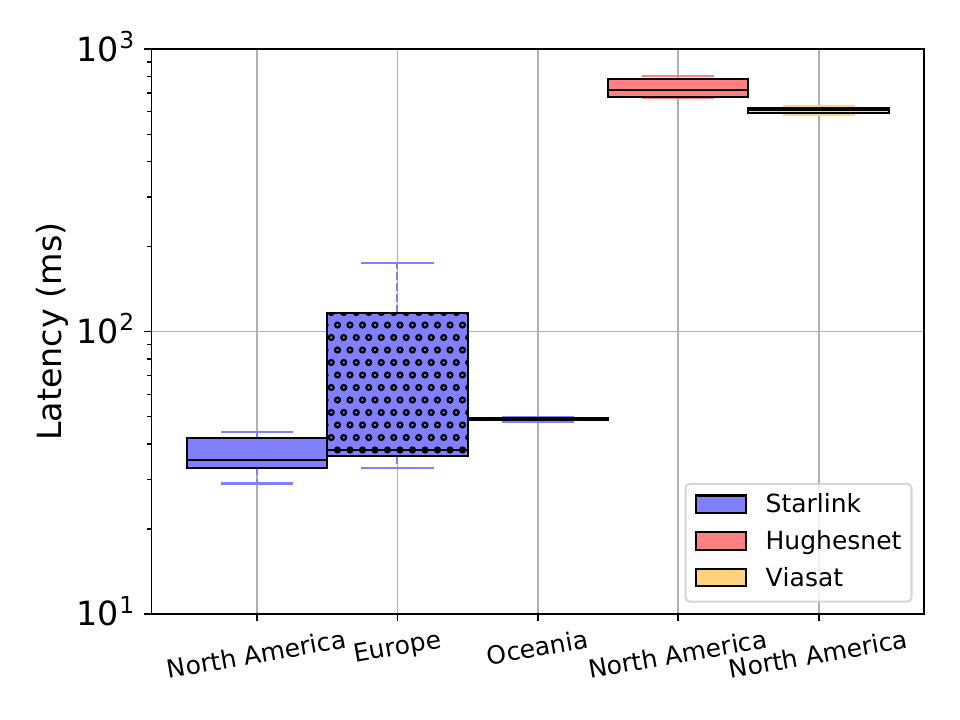}\label{fig:prolific:speedtest:latency}}\hfill
    \vspace{-0.1in}
    \caption{Analysis of speedtest data (\textit{fast.com})  organized by continent (North America, Europe, Oceania) and satellite-based operator (Starlink, HughesNet, Viasat).}
    \vspace{-0.25in}
\label{fig:prolific:speed}
\end{figure*}
In our data collection, we instrument the addon to load two demo pages provided by Akamai to compare the performance of HTTP/1.1 (H1)~\cite{akamaiH1} and HTTP/2 (H2)~\cite{akamaiH2}. Our rationale is threefold. First, these pages have constant sizes, no impact of personalized ads, and they bypass the browser cache using unique Web object identifiers. Second, these pages are hosted by a large CDN with a wide footprint and thus at a relatively short distance to all our users.%
Third, they are \textit{designed} to allow a comparison of the performance of H1 and H2. 

The webpage test is followed by video streaming, where a predefined YouTube video is played for 60 seconds. During video playback, injected JS code monitors the YouTube's ``Stats-for-Nerds'' which reports statistics like video quality, data available in the buffer, and number of dropped frames. To avoid interference with the browser cache for returning testers, we use a 10 hours 4K video and increase the starting playback time by one hour, \eg by appending \texttt{\&t=3600s} to the URL. %

\pb{Data collection.} During April-May 2023, we launched  Prolific measurement campaigns targeting the 56 testers we previously identified as SNO subscribers. Over one month, we recruited 20 testers (10 on Starlink, 5 on HughesNet, and 5 on Viasat) willing to install and run our addon once a week (\ie four runs in total) on random days and times. The remaining 36 Prolific testers did not accept our jobs. Some had temporarily no access to their SNO, others were not willing to install software on their machines, and the rest just never responded to our job requests and direct messages.

\subsection{Results}
\pb{Speedtest.} Figure~\ref{fig:prolific:speed} summarizes the speedtest analysis per SNO (Starlink, HughesNet, and Viasat), and per continent (North America, Europe, and Oceania). We report on both download and upload speeds, and network latency measured between each tester and the speedtest service provided by \textit{fast.com}. Overall, Starlink offers much higher speeds in both download (70-150~Mbps) and upload (6-21~Mbs). While Viasat and HughesNet achieve comparable upload speeds (3~Mbps), Viasat offers much higher download speeds (10-40~Mbs), in line with what is advertised~\cite{hughes_vs_viasat}, while HughesNet testers never experience more than 3~Mbps, which is far from the download speed advertised by this provider (25~Mbps~\cite{hughes_vs_viasat}). With respect to the per-continent analysis, Starlink testers reach similar download speeds ($\sim$80~Mbps) in both North America (nine testers) and Oceania (only one tester located in New Zealand). Much higher speeds (median of 150~Mbps) are instead reached by five European testers (Italy, UK, Netherlands, and Czech Republic). A similar trend is observed for upload speeds as well, with European testers reaching the fastest speeds ($\sim$21~Mbps), followed by New Zealand with a stable speed of 13~Mbs, and North America with more variable speeds and a much lower median upload speed of 6~Mbps.

\begin{figure*}[tb]
    \subfloat[\texttt{Download time of \texttt{jquery.min.js} across CDNs.}]{\includegraphics[width=0.32\linewidth]{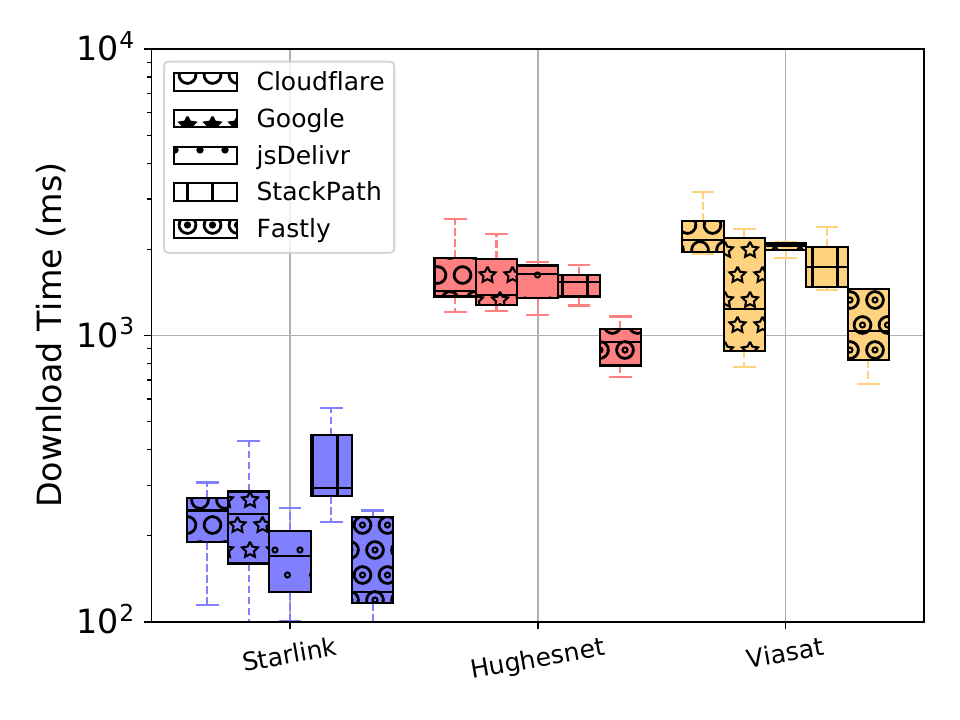}\label{fig:prolific:web:CDN}}\hfill
    \subfloat[\texttt{PLT of Akamai's demo page when using H1 and H2.}]{\includegraphics[width=0.32\linewidth]{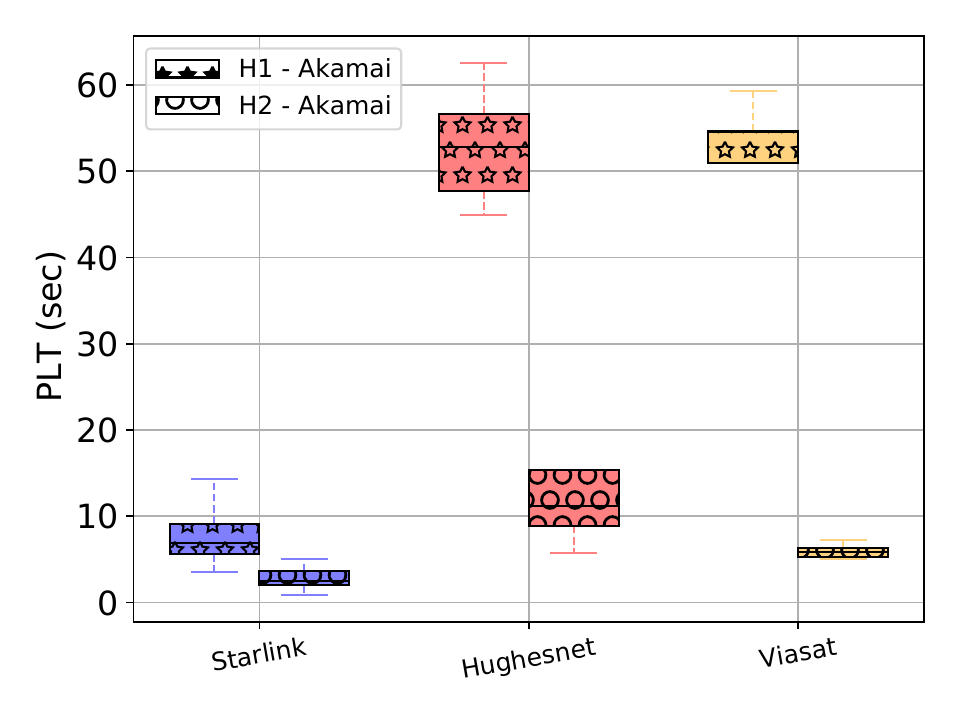}\label{fig:prolific:web:TFB}}\hfill
    \subfloat[\texttt{CDF of DNS lookup times. Starlink uses Cloudflare DNS.}]{\includegraphics[width=0.32\linewidth]{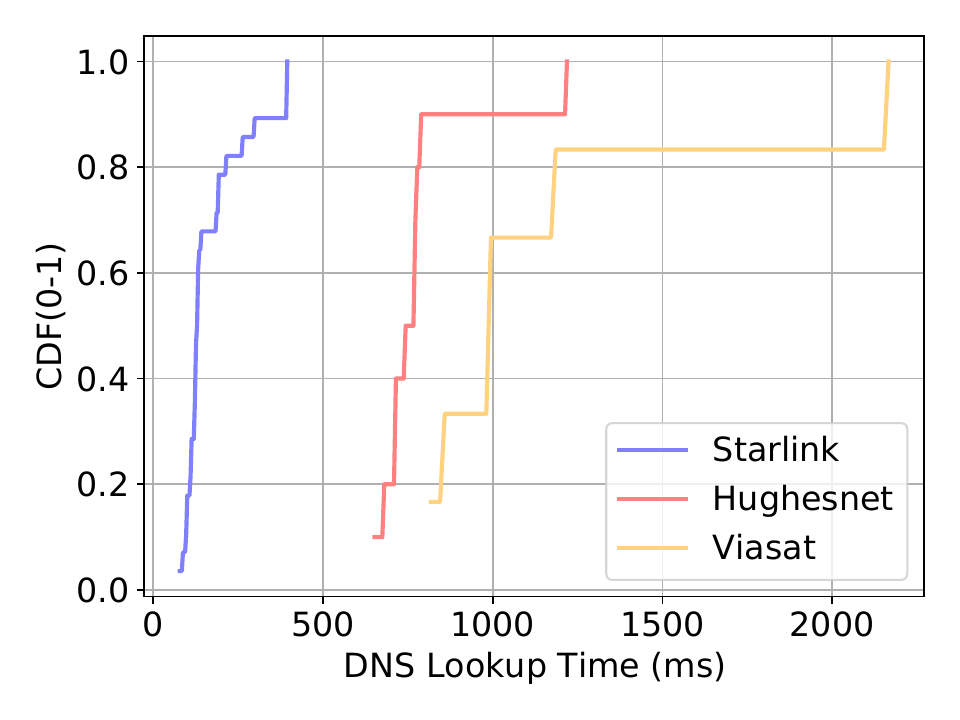}\label{fig:prolific:web:DNS}}\hfill
    \vspace{-0.1in}
    \caption{Web browsing analysis: Starlink vs.~HughesNet vs.~Viasat.}

\label{fig:prolific:web}
\end{figure*}

We now focus on the RTT between Prolific testers and  \textit{fast.com} servers. Figure~\ref{fig:prolific:speedtest:latency} shows that, despite Viasat and HughesNet both being GEO networks, Viasat manages to save about 100~ms, \ie median latency to \textit{fast.com} of 600~ms compared to 720~ms with HughesNet. Starlink testers benefit from the low latency offered by LEO satellites, with median latencies of 35.0~ms (North America), 38.0~ms (Europe) and 49~ms (New Zealand). These latencies are comparable with the RTT measured between RIPE probes and Starlink PoPs (see Figure~\ref{fig:ripe_global_first_hop_RTT} and~\ref{fig:rip_states_cdf_groundstation}), indicating that \textit{fast.com} servers are likely co-located with these PoPs. Note that the high latency values measured in Europe (up to 150~ms) are all associated with runs from the London tester. Given the results from  Figure~\ref{fig:ripe_global_first_hop_RTT}, this latency represents an outlier due to some issues on the test side, \eg bad WiFi setup. 

\pb{Web browsing.} We now focus on measurements related to Web browsing. Specifically, we report the time taken to download the popular \texttt{jquery} library from multiple CDNs, the time to perform DNS lookups, and how quickly Akamai test pages are loaded using both HTTP/1.1 (H1) and HTTP/2 (H2). Figure~\ref{fig:prolific:web:CDN} shows, per SNO, boxplots of download time of jquery's minified version (\texttt{jquery.min.js}) via five CDNs:  Cloudflare, Google, jsDelivr, StackPath, and Fastly. For all three SNOs, Fastly provides the fastest downloads, with a median of 127~ms (Starlink), 950~ms (HughesNet), and 1,036~ms (Viasat). When focusing on Starlink, jsDeliver is the second fastest CDN with a median download time of 170~ms; when we analyze the response headers, we find that all jsDeliver requests are handled by Fastly. This result suggests that jsDeliver is effectively identifying the best performing CDN; however, this process requires one extra RTT which eliminates the benefit for GEO SNOs. For example, jsDeliver adds an extra 700~ms (download time of 1,641~ms) for HughesNet, making it slower than Cloudlfare (1,427~ms), Google (1,385~ms), and StackPath (1,537~ms).

Although not shown to avoid cluttering the figure, a similar trend can be observed for \texttt{jquery.js}, \ie the non-minified version with a size of about 87~KB (versus 32~KB). In this case, Fastly serves the file in 190.0~ms (Starlink), 1,450~ms (Viasat), and 1,620~ms (HughesNet). Given the impact of JS libraries like jquery to start rendering webpages, this result showcases the importance of careful webpage development when dealing with testers behind such high latencies. A similar observation holds also for Figure~\ref{fig:prolific:web:TFB} which reports the ``page load time'' of Akamai's demo page when comparing H1 and H2.  Adopting H2 is paramount for GEO testers, enabling performance comparable to using H1 on Starlink. Notice also the compound effect of the lower latency provided by Viasat (about 100~ms according to \textit{fast.com}, see Figure~\ref{fig:prolific:speedtest:latency}) when loading complex pages (hundreds of small objects in the case of this demo) which can speed up webpage loads by multiple seconds. 

Finally, Figure~\ref{fig:prolific:web:DNS} shows the CDF of the DNS lookup time across SNOs. For this analysis, we leverage both the data collected by the addon when loading Akamai demo pages, and data collected by our server when loading the census form (see Section~\ref{sec:meth:prolific}). We have verified, using \textit{ https://test.nextdns.io/} as discussed in Section~\ref{app:meth}, that all Prolific testers rely on what provided by their SNO, \ie Cloudflare for Starlink and their own DNS service for both both Viasat and HughesNet. Note that while DNS entries can be cached along the path, \ie  browser, OS, router, and resolver, we minimize this issue by relying on unpopular domains (Akamai demo and our own domain) with short TTL (300 seconds). We further filter DNS lookup times smaller than the minimum RTT measured for each SNO. Figure~\ref{fig:prolific:web:DNS} shows median DNS lookup times of 130~ms (Starlink), 755~ms (HughesNet), and 985~ms (Viasat). Given the overall faster RTT offered by Viasat, this result suggests that HughesNet-provided DNS is faster than Viasat, and that its testers rely on default setting, \ie they did not manually set DNS in their devices. We would otherwise expect faster DNS lookup times for Viasat if a cloud-based DNS service (\eg Google or Cloudflare) was used, since in this case the lower RTT offered by Viasat would dominate. We further verified this claim by messaging HughesNet and Viasat testers on Prolific.

\begin{figure*}[tb]
    \subfloat[\texttt{Download speed.}]{\includegraphics[width=0.31\linewidth]{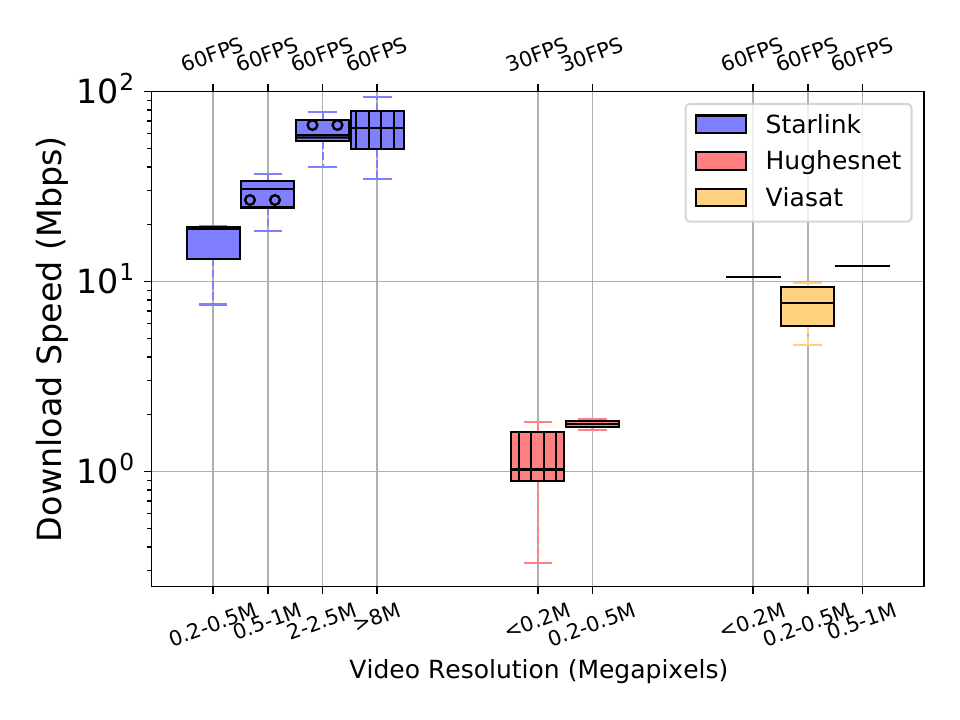}\label{fig:prolific:yt:rate}}\hfill
    \subfloat[\texttt{Buffer health.}]{\includegraphics[width=0.31\linewidth]{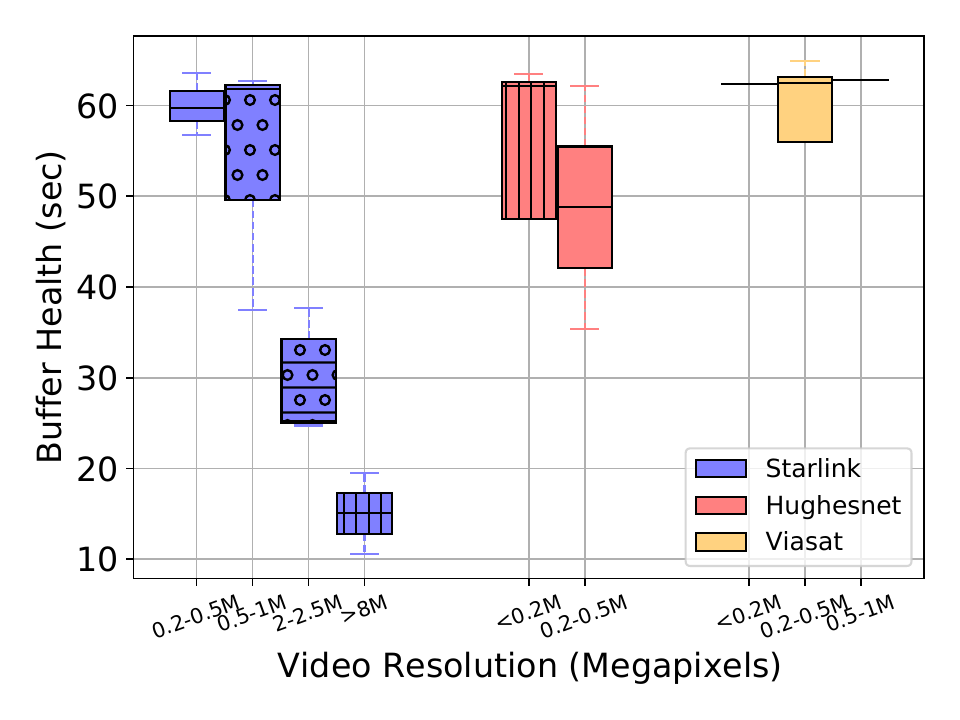}\label{fig:prolific:yt:buffer}}\hfill
    \subfloat[\texttt{Dropped frames.}]{\includegraphics[width=0.31\linewidth]{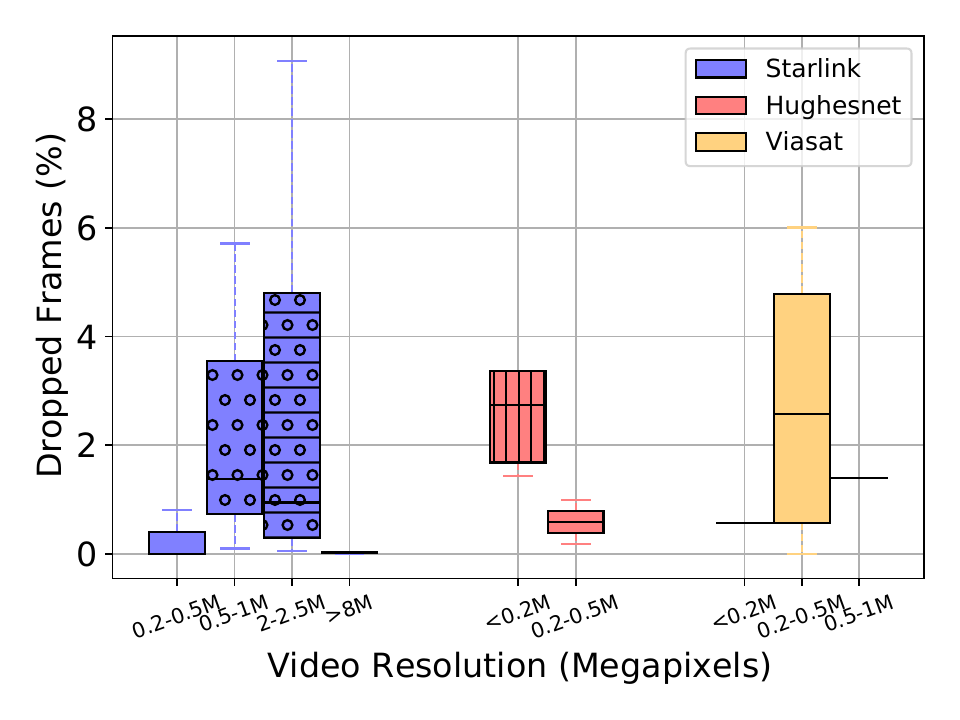}\label{fig:prolific:yt:dropped}}\hfill
    \vspace{-0.1in}
    \caption{YouTube analysis: Starlink vs.~HughesNet vs.~Viasat.}
\label{fig:prolific:youtube}
\end{figure*}

\pb{Video streaming.} Figure~\ref{fig:prolific:youtube} summarizes the video streaming analysis per SNO reporting three metrics (download speed, buffer health, and percentage of frames dropped) as a function of the video quality experienced. We express the video quality as \textit{megapixels} derived as ``width'' multiplied by ``height'' from the video resolution. For example, a 1080p video (1920x1080) corresponds to about two megapixels. This step is necessary since we observe 18 unique video resolutions, which would make the figure unreadable. Further, we report the median video quality for each experiment since the YouTube player is adaptive to network conditions. Note that the test video has a maximum quality of 2160p (3840x2160) or about 8 megapixels. 

Overall, Figure~\ref{fig:prolific:youtube} shows that only Starlink testers are able to play the video at high resolution (2 megapixels or higher). On the other hand, for both HughesNet and Viasat, about 0.5 megapixels (\ie less than the popular 360p, or 480x360) is the most common resolution experienced by their testers, although Viasat testers manage to maintain 60 frames-per-second (FPS) and sometime venture in higher resolutions of about 1 megapixel. This is due to a lack of download bandwidth as shown in Figure~\ref{fig:prolific:yt:rate} (download speed measured by YouTube) which confirms the trend observed in Figure~\ref{fig:prolific:speedtest:down}. Indeed, 1080p or higher is hard to achieve also for Starlink testers, requiring sacrificing \textit{buffer health}, \ie the amount of seconds of video available to be played. 

Figure~\ref{fig:prolific:yt:buffer} shows how most runs, regardless of the SNO, are characterized by buffer health of between 40 and 65 seconds, whereas this number drops to 15 and 30 seconds when considering high resolutions for Starlink testers. Conversely, the percentage of frames lost is less correlated with the video quality, \eg no frame was lost for Starlink runs where the video was played at full resolution (4K or more than 8 megapixels). These overall high losses rate are more likely due to the frequent handovers (for Starlink) as previously discussed in Section~\ref{sec:bird}. \camfix{Note that YouTube does not directly report the number of video stalls detected. However, this metric can be approximated by the likelihood of the buffer health to reach zero. Out of 56 testers, only 4 testers (2 on Starlink and one on Viasat and HughesNet) have experienced some video stalls, between a minimum of 5\% and up to 32\% of the video playback.}\reviewfix{F7}

\section{Conclusion}

By \textit{opportunistically} utilizing a wide variety of publicly available data sources together with  user studies, this paper performs what we believe is the first study which compares the performance of all three kinds of SNOs (LEO, MEO and GEO) and identifies factors critical for their performance. \camfix{We observe that LEO networks experience greater relative variations in jitter, which can impact applications that rely on a consistent latency profile. Furthermore, we identify a number of potential factors that web applications can optimise, such as the choice of CDN used etc., as well as factors under SNO control, such as the location and geographic distribution of PoPs, which can each help improve global performance of satellite-based networks. We believe that these findings will be valuable to various stakeholders, from application developers looking to design their products with next-generation satellite access in mind, to CDN and satellite network operators seeking to make careful technology selection (for example, choosing the right locations to peer with SNOs). The study also highlights the benefits of bringing together different public data sources (M-Lab, RIPE, CAIDA) to gain a comprehensive understanding of connectivity performance in satellite networks. }\reviewfix{B2, E9, A4}

\camfix{As a part of future work, we plan to investigate temporal trends and assess the underlying causes of latency fluctuations over time among different operators. Additionally, we plan to conduct a more in-depth analysis of TCP traces to thoroughly examine retransmission rates and study the performance characteristics of satellite internet service providers.}\reviewfix{B2, C1, C3, A5}

\section*{Acknowledgments}
This work is supported by EU Horizon Framework grant agreement 101060294 (XGain)  and the ISOC Pulse Research Fellowship awarded to Aravindh Raman.

\bibliographystyle{plain}
\bibliography{main}

\appendix
\section*{Appendix}
\section{Ethics}
\label{sec:appendix:ethics}
Our work uses the performance traces from \one two widely-used speed test measurement suites and testbeds (operated by M-Lab and RIPE) and \two human participants through Prolific. The data from former sources are anonymized \cite{measurementlabprivacy, ripeatlaslegal} and are publicly available (through Google BigQuery and APIs). In fact,  M-Lab recommends to use the data for research by waiving the copyright and related rights to the data. In case of latter, Prolific anonymises all the participant information~\cite{prolificdataprotection}. There are no methods available to decode any Personal Identifying Information from Prolific. We also contacted the Institutional Review Board (IRB) office at our University and they deemed that this was not human subjects research%
 and as the tool is mainly automated to run and collect results.

\section{Exhaustive List of SNOs}
\label{sec:appendix:snos}
Table~\ref{tab:operators} shows a curated list of 67 ASNs belonging to 41 SNOs we have identified using the methodology we developed in this paper, specifically ``ASN-to-SNO Mapping'' described in Section~\ref{sec:meth:identify}. %

\begin{table*}[h]

\begin{tabular}{l|l}
\hline
SNO              &      ASN    \\
\hline
arqiva              &    15641 \\
avanti              &    39356 \\
awv                 &    46869 \\
colinanet           &   262168 \\
comsat              &    36614 \\
comsat (png)        &   136940 \\
comtech             &   394318 \\
elara               &   262927 \\
eutelsat            &   204276 \\
            &    34444 \\
            &    15829 \\
globalsat           &    28503 \\
gravity             &   131202 \\
hellas-sat          &    41697 \\
hughes              &    28613 \\
              &     1358 \\
              &    63062 \\
              &    12440 \\
              &    44795 \\
              &     6621 \\
intelsat            &    26243 \\
            &    46982 \\
io                  &    17411 \\

\hline
              \end{tabular}
\hfill
\begin{tabular}{l|l}
\hline
SNO              &      ASN    \\
\hline
isotropic           &    36426 \\
kacific             &   135409 \\

kvh                 &    25687 \\
lepton (kymeta)     &    20304 \\
linkexpress         &    20660 \\
marlink             &     5377 \\
             &    44933 \\
             &    55784 \\
             &     8841 \\
             &   210314 \\
             &     8264 \\
             &    37101 \\

maxar               &   393938 \\
navarino            &   203101 \\
netsat              &   133933 \\
network innovations &     1821 \\
nomad global        &   395786 \\

o3b                 &    60725 \\
oneweb              &      800 \\
panasonic           &    64294 \\
ses                 &   201554 \\
                 &    12684 \\
\hline
\end{tabular}
\hfill
\begin{tabular}{l|l}
\hline
SNO              &      ASN    \\
\hline

sound \& cellular    &    63215 \\
speedcast           &    38456 \\
ssi                 &    22684 \\
starlink            &    14593 \\
            &    27277 \\
telalaska           &    10538 \\
telesat             &    19036 \\
televera            &   265515 \\
thaicom             &    63951 \\
ultisat             &   393439 \\
viasat              &    13955 \\
              &    25222 \\
              &    46536 \\
              &    18570 \\
              &    16491 \\
              &    40306 \\
              &     7155 \\
              &    40310 \\
              &    40311 \\
              &    23354 \\
              &    31515 \\
worldlink           &    11902 \\

\hline
\end{tabular}
\caption{Curated list of SNO-ASN mapping captured from ASdb and manually visiting each operator website.}
\label{tab:operators}
\end{table*}

\section{BGP Peering Analysis of SNOs}

\begin{figure*}[h]
    \includegraphics[width=1.0\linewidth, trim=0mm 10mm 0mm 5mm, clip=true]{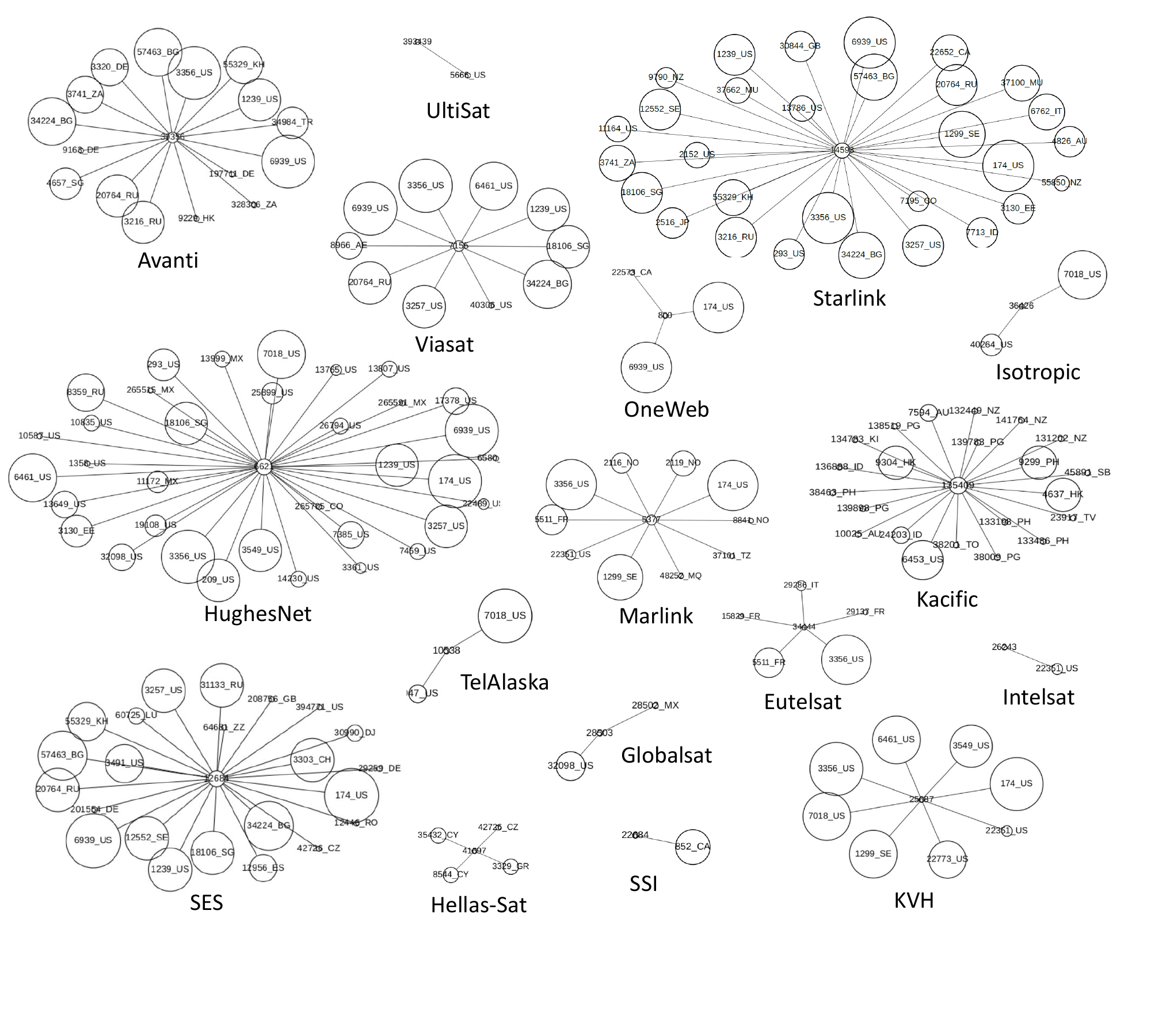}
    \vspace{-4ex}
    \caption{BGP peering visualization of SNOs (based on BGP route-views collected on 2023/1/1).}
\vspace{-0.1in}
\label{fig:bgp_viz}
\end{figure*}

\begin{figure*}[tb]
    \subfloat[Starlink: its peering has evolved significantly across the globe.]{\includegraphics[width=1.0\linewidth,trim=0mm 3mm 0mm 3mm,clip=true]{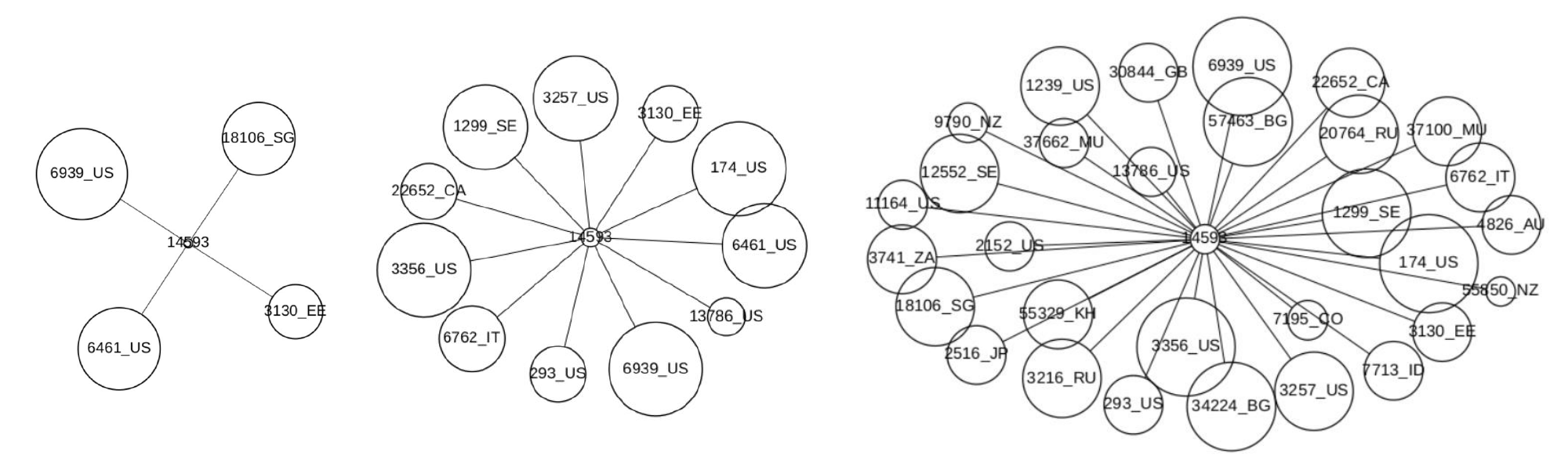}}\label{fig:bgp:starlink}\hfill
    \subfloat[HughesNet: its peering has remained the same.]{\includegraphics[width=1.0\linewidth,trim=0mm 1mm 0mm 3mm,clip=true]{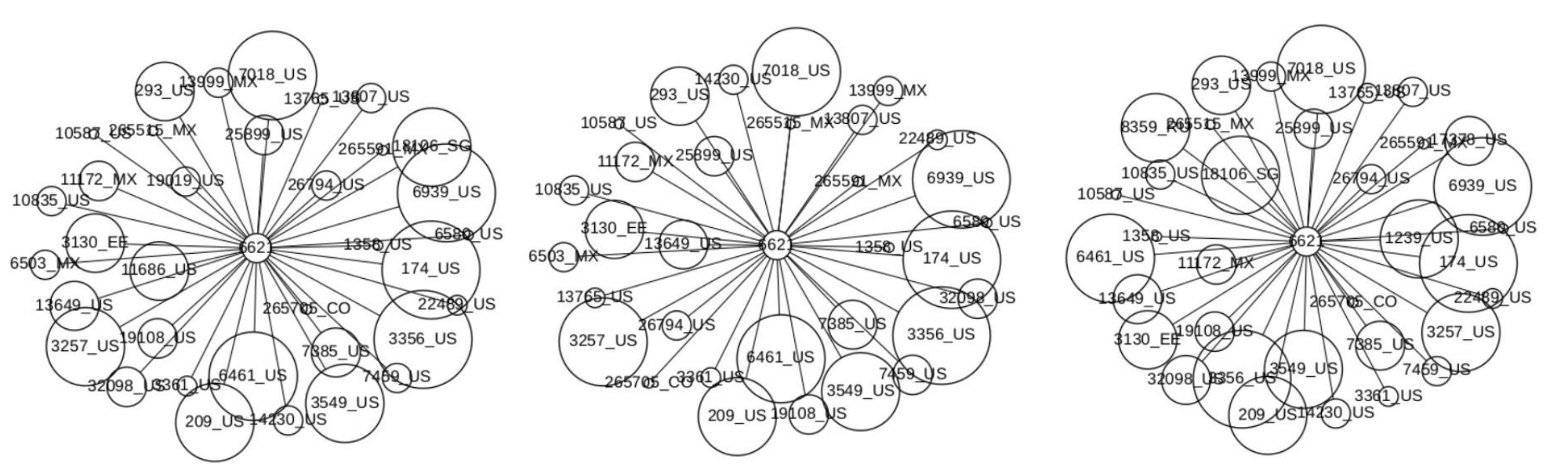}}\label{fig:bgp:hughes}\hfill
    \subfloat[Viasat: its peering has expanded from the US to non-US regions worldwide.]{\includegraphics[width=1.0\linewidth,trim=0mm 10mm 0mm 10mm,clip=true]{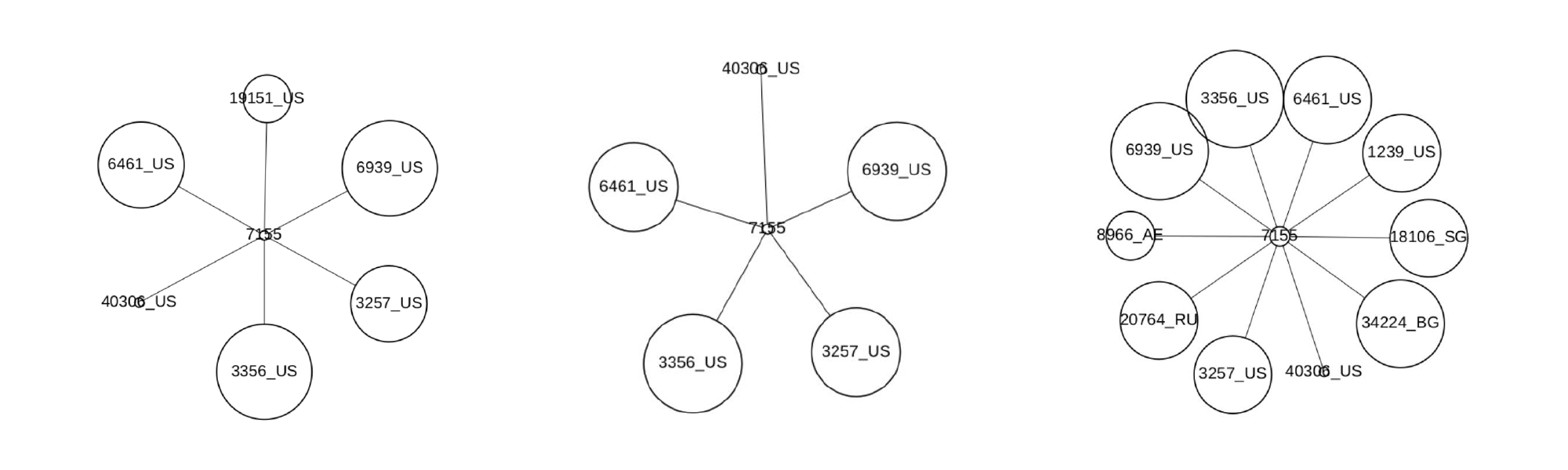}}\label{fig:bgp:viasat}\hfill
    \subfloat[Marlink: its one tier-1 provider in the US changed from Level3 (3549) to Cogent (174).]{\includegraphics[width=1.0\linewidth,trim=0mm 10mm 0mm 10mm,clip=true]{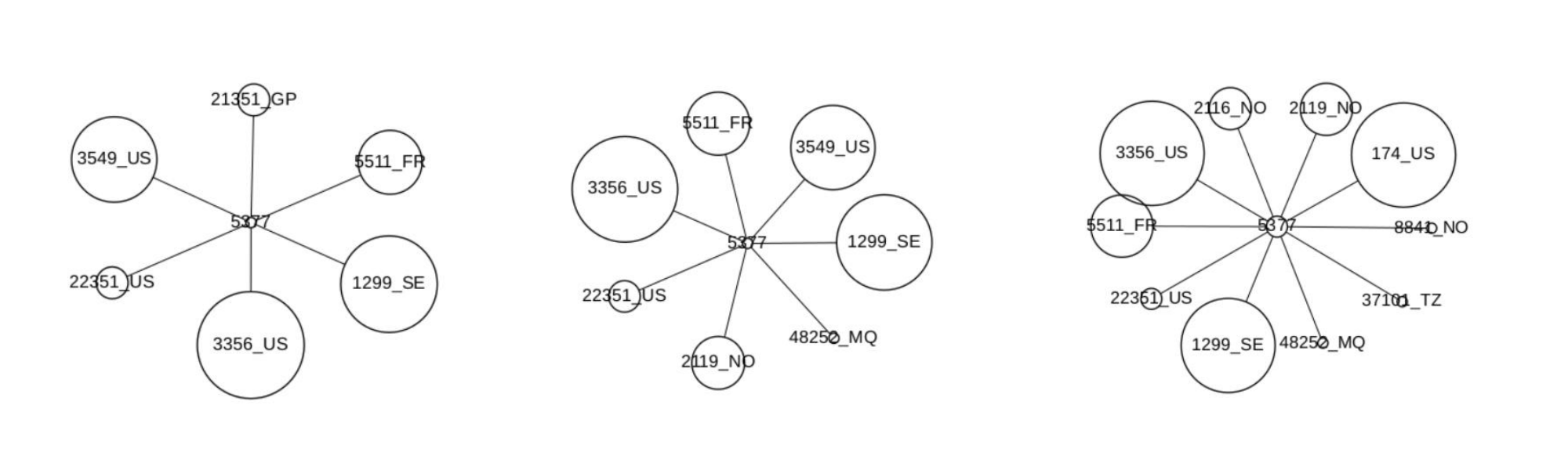}}\label{fig:bgp:marlink}\hfill
    \caption{Historical evolution of BGP peerings of SNOs: 2021/1 on left, 2022/1 on center, and 2023/1 on right.}
\label{fig:bgphistory}
\end{figure*}

Our BGP-peering-based PoP characterization described in Section~\ref{bgp_analysis} allows us to examine and compare different SNOs' ground infrastructures as well as their historical evolution over time.  Figure~\ref{fig:bgp_viz} shows BGP peering visualization of the SNOs evaluated in Figure~\ref{fig:sno:retx}, as captured by BGP route-views collected in January, 2023. Figure~\ref{fig:bgphistory} compares representative growth patterns of several SNOs. For this analysis, we use BGP route-views snapshots captured from three separate periods with one year interval (from 2021/1 to 2023/1).

\clearpage
\section{Prolific Census Results}
\begin{figure}[!htb]
    \center
    \includegraphics[width = 0.8\linewidth]{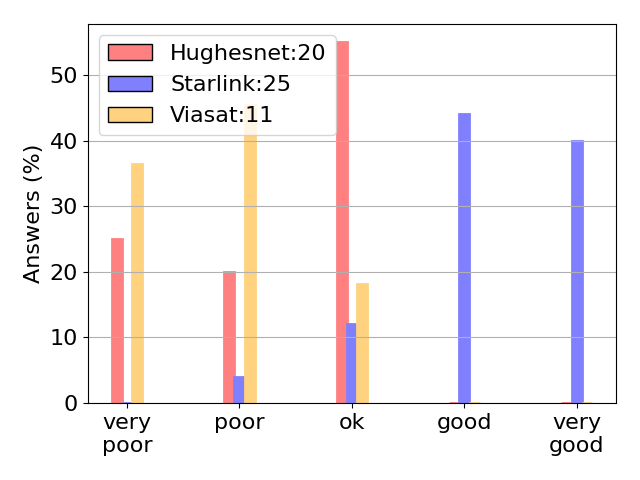} 
    \vspace{-10pt}
    \caption{Feedback from 56 Prolific testers on the quality of their SNO. Scores  from 1 (very poor) to 5 (very good). Data collected in March 2023.}
    \label{fig:census:score}
\end{figure}

Figure~\ref{fig:census:score} visualizes feedback (from ``very poor'' to ``very good'') from 56 Prolific testers who are also subscribers of HughesNet, Starlink, and Viasat. 

\clearpage

\end{document}